\documentclass[conference]{IEEEtran}
\usepackage{hyperref}
\usepackage[hyphenbreaks]{breakurl}
\usepackage{graphicx}
\usepackage{subfigure}
\usepackage{algorithmic}
\usepackage{algorithm}
\usepackage{textcomp}
\usepackage{xcolor}
\usepackage{amsmath,amssymb,amsfonts}
\usepackage{lineno,hyperref}
\def\BibTeX{{\rm B\kern-.05em{\sc i\kern-.025em b}\kern-.08em
    T\kern-.1667em\lower.7ex\hbox{E}\kern-.125emX}}
\begin{document}

\title{Allocating Limited Resources to Protect a Massive Number of Targets using a Game Theoretic Model}

\author{\centering
\IEEEauthorblockN{Xu Liu, Xiaoqiang Di$^*$, Jinqing Li, Huan Wang, Jianping Zhao, Huamin Yang, Ligang Cong}
\IEEEauthorblockA{\textit{School of Computer Science and Technology, Changchun University of Science and Technology, Changchun, China} \\
\textit{Jilin Province Key Laboratory of Network and Information Security, Changchun, China}\\
$^*$Corresponding author:  dixiaoqiang@cust.edu.cn}
\IEEEauthorblockN{ Yuming Jiang}
\IEEEauthorblockA{\textit{Department of Information Security and Communication Technology} \\
\textit{Norwegian University of Science and Technology, Trondheim, Norway}}
}

\maketitle

%
%
%

\begin{abstract}
Resource allocation is the process of optimizing the rare resources. In the area of security, how to allocate limited resources to protect a massive number of targets is especially challenging. This paper addresses this resource allocation issue by constructing a game theoretic model. A defender and an attacker are players and the interaction is formulated as a trade-off between protecting targets and consuming resources. The action cost which is a necessary role of consuming resource, is considered in the proposed model. Additionally, a bounded rational behavior model (Quantal Response, QR), which simulates a human attacker of the adversarial nature, is introduced to improve the proposed model. To validate the proposed model, we compare the different utility functions and resource allocation strategies. The comparison results suggest that the proposed resource allocation strategy performs better than others in the perspective of utility and resource effectiveness.
\end{abstract}




\section{Introduction}\label{sec:introduction}
Resource allocation has always been a complex problem, especially when driven by security requirements. How to devise a mechanism to control the trade-off between the cost of protection and the achieved security utility is an open challenge \cite{cloudissue2014}. In the AWS re:Invent 2014, the AWS engineer claimed that Amazon had nearly 28 total sets across the world, each of which has one or more data centers with a typical facility containing 50,000 to 80,000 servers \cite{vmNum}. To protect these servers against attack and maintain their consistent operation, cloud providers will implement security strategy. For example, they can protect targets (eg. virtual machines, VMs) by setting up resource reservations to analyze the operation of targets and then respond the attack quickly, which is followed by a lot of resource consumption \cite{url1}. Therefore, a trade-off problem could be abstracted between consuming resources and protecting targets. Especially, when the number of available resources or resource budget is fixed and limited for all the targets, how to allocate limited resources to protect a massive number of targets is a vital issue in the security area.

The extreme approach may be to allocate security resources to cover all the targets \cite{all}. For instance, setting up the full resource reservations for all the VMs, which will lead to almost double resource consumption. The common approach may be to protect those targets with the most value \cite{alert}. For instance, setting up the resource reservation for the VMs that store the most data or the sensitive data (eg. financial data). The former approach fails to consider resource constraints and effectiveness, however, the available resources may not be sufficient to protect all the targets on the one hand, on the other hand, resources allocated to some empty targets may be inefficient. The latter approach does not account for the adversarial nature and perspective-taking of the attacker. An attacker who can learn about a defender's possible target protection strategy can exploit this knowledge to launch an attack on the targets that the defender does not protect.

This paper focuses on developing a general resource allocation method to address the trade-off between security gain and resource consumption. The goal is to resolve the problem of how to utilize limited resources to efficiently protect massive targets against attack. How to build a mathematical model to describe this problem is the key. For example: (1) How to maintain security while allocating resources? (2) How to simulate an attacker of the adversarial nature?

In the previous studies \cite{alert,securityresource,lim3,lim6,reward,16}, the number of allocated resources is measured by defense probability. But the importance and emphasis of resource allocation weakens in such scheme. In general, performing different actions on a target will result in different outcomes. If an action is successful, the actor will obtain some benefit as a reward; otherwise, the actor will lose some assets as a penalty. No matter whether an action is successful, the actor will incur some cost by taking the action. Recent studies about the effort of deterrence and risk preferences in the security games \cite{Budget2017,Risk2017} have analyzed the impact of risk preference on the defense effort and deterrence level, and the impact of defender's cost on the investment strategy. Meanwhile, statistics show that a large data center costs between \$10 million and \$25 million per year and the corresponding maintenance costs account for nearly 80\% of its total cost \cite{MTence}. So it's clear that action cost is an important factor which cannot be ignored. By combining the rewards, penalties, costs and probabilities of actions in some manner, it may be possible to describe our problem.

Game theory, an important tool for analyzing real-world resource allocation problems, such as the assignment of cyber analysts \cite{alert} and patrolling strategies \cite{patrolling08,patrolling09}, provides an alternative solution. However, in most of the previous studies \cite{securityresource, lim3, lim6, reward} on game theoretic resource allocation, only the reward and penalty associated with an action have been included in the game utility function, but the action cost has been ignored. In the real world, no matter what one wants to do, an action cost is often necessary. This cost might be measured in monetary units, physical resources, abstract resources and so on. Whatever it is, it can be abstracted as a mathematical expression. Hence, we include cost additionally in the Stackelberg game \cite{11} utility function, and analyze the impact of different parameter value configurations on the defender's utility.

Since both the defender and attacker are intelligent and have the perspective-taking ability, we consider an interaction in which the defender designs a resource allocation strategy first and the attacker subsequently develops an attack strategy. Although the attacker has the ability to consider the situation from the perspective of the defender, the attacker might also take abrupt actions that lie outside the defender's expectations. This type of attacker, who is of the adversarial nature, can be simulated by the quantal response (QR) model, which has received widespread support in the literature on modeling human behavior in games \cite{20}. In this paper, we introduce it into the proposed Game Theoretic Resource Allocation (GTRA) model to simulate adversarial reasoning.

The efficient resource allocation strategy for the defender is obtained from an optimization algorithm. Three indicators, namely vulnerability, coverage and effectiveness, are designed to evaluate the effectiveness of our strategy. We compare the equilibrium strategy based on the proposed GTRA with the one based on a game utility function without considering the action cost. And also compare with four extreme resource allocation strategies, namely average allocation strategy, partial allocation strategy, random allocation strategy and full-coverage strategy. The experimental results demonstrate the effectiveness of our proposed GTRA model.

The contributions of this paper can be summarized as follows:

\begin{itemize}
\item[(1)] To emphasize the action cost in resource consumption. The players' action costs are included in the game utility function as an independent item. The numerical analyses prove that this type of resource measurement can improve the utility and effectiveness. 

\item[(2)] To better balance target security and resource consumption. The obtained Nash equilibrium strategy is selected as the defender's resource allocation strategy because it outperforms the other extreme resource allocation strategies in terms of both security and effectiveness.

\item[(3)] The constructed GTRA model provides advice based on the target parameters to assist in determining the appropriate quantity of resources to protect a massive number of targets. 
\end{itemize}

The remainder of this paper is organized as follows. Section \ref{sec:related} and section \ref{sec:problem} describe the related work on resource allocation and our problem, respectively. Game theoretic model, QR model and the proposed algorithm are presented in section \ref{sec:model}. The numerical analyses are discussed in section \ref{sec:experiment}. The final section summarizes the paper and outlines directions for future work.

\section{Related Work}\label{sec:related}
Resource allocation is defined as the economical distribution of resources among competing groups of people or programs \cite{cloudissue2014}. Game theory has been applied in resource allocation to better capture the interaction between resource provider and user, and show the economic nature of resource allocation. The previous studies can be roughly classified into two categories based on the different participants considered: security-driven resource allocation between a resource provider and an attacker; and demand-driven resource allocation between a resource provider and a legitimate user.

Demand-driven resource allocation can be further subdivided into cost-scheme-based, performance-scheme-based and mixed-scheme-based resource allocation. The original pricing scheme is used for the allocation of resources of a single type, such as bandwidth \cite{FairBandwidth, CloudBandwidth, DatacenterBandwidth}, offload \cite{Offload,OffloadScheduling}, or cache \cite{CacheAllocation}. With the development of the Internet, resource provider could provide nearly all the resources that users need, such as cloud computing provider provides on-demand resources including storage, memory, bandwidth and so on. Multi-resource pricing schemes \cite{MultiAllocation}, such as the cost-optimized scheme considering multiple resources \cite{Cost-Optimized}, have emerged. Meanwhile, since user requests are becoming necessary while providing service, some research has focused on user-demand-driven resource allocation \cite{auctionSLA,PerfAllocation,qosResource}. Later, the cost-optimized and performance-based schemes are combined to allow a resource provider to achieve a win-win objective in which resource provider obtains the maximum profit while the user receives the best experience \cite{optimizing_satisfiability,MTence,PerfPricing,Liu2017Analysis}.

However, during the pursuit of the best experience and the maximum benefit, security issues increase, and security-driven resource allocation become a research hotspot, especially when resources are limited and cannot cover all the targets that require protection. The American institute Teamcore conducted a project with the theme of "AI and game theory for public safety and security", and their achievements have been applied in various areas. ARMOR \cite{22} was deployed to develop randomized checkpoints and a patrol route strategy at Los Angeles International Airport. GUARDS \cite{23} was developed to assist airports in allocating limited air police resources to protect more than 400 United States airports. Federal Air Marshals used IRIS \cite{28} to provide scheduling coverage for potential attacks. PROTECT \cite{27} was deployed to generate randomized patrolling schedules for the US Coast Guard. These cases are typical instances of limited security resource allocation using game theoretic model.

Other game theoretic studies have also produced good results. One study \cite{alert} investigated an intelligent allocation method for assigning limited cyber analysts to analyze a massive number of security alerts in a network. Another work \cite{16} developed new models and algorithms that could scale to highly complex instances of limited security resource allocation games. Their new methods performed faster than known algorithms when solving massive security games. In further research \cite{securityresource} based on a previous work \cite{lim3}, efficient algorithms were developed to compute the best responses of security forces to different adversary models when resources are limited, and it was proven that the proposed response strategy was superior because it relaxed the assumption of perfect rationality. An additional study \cite{lim6} proposed a game theoretic scheme for developing dynamic and randomized security strategies that consider an adversary's surveillance capabilities. The experimental results showed that the proposed algorithm outperformed the existing approaches. 

Although these works have utilized the nature and principles of game theory to determine optimal resource allocation strategies, most of them considered only rewards and penalties in their allocation strategies. Recent works \cite{Budget2017,Risk2017} specially examined the effect of risk preferences on deterrence, and analyzed the impact of the defender's cost on its investment, which demonstrated that the cost of actions cannot be ignored. Nonetheless, in the previous works \cite{alert,securityresource,lim3,lim6,reward,16}, the action cost was measured by defense probability simply, which inclines to analyze the impact of defense instead of action cost. Therefore, the game theoretic approach that includes the action cost independently is required to perform the resource allocation in the security area. 

\section{Problem Description}\label{sec:problem}
This paper considers a common scenario of a defender and an attacker. The defender's responsibility is to protect the security of $N$ targets using $M$ resources, so it allocates resources to targets as its action. By contrast, the attacker's intention is to attack the targets, and such attack also costs resources. For both sides, the benefit of consuming resources can be measured in terms of the security gain. The resources can be computing, storage, energy or even monetary units, and the security gain indicates the return of protecting the targets by consuming resource. Although the units of resources and returns are different, they can be abstracted into the numerical value by mathematical methods. In this paper, we put emphasis on analyzing the relationship between them by setting various parameter configurations to simulate the different scenarios. For example, if the defender allocates resources to a target $i$, this target will be relatively more secure than a target without being covered by resources, which can be configured with a bigger security gain. 

Therefore, the defender obtains a security gain by expending resources, which can be abstracted as a limited resource allocation problem, that is, the problem of how the defender should allocate $M$ resources to protect $N$ when $M$ is far less than $N$. The defender wants to achieve the greatest security gain while minimizing resource consumption. Therefore, this is a trade-off problem between protecting targets and consuming resources. Table \ref{tab1} lists the parameters used in this paper. $T=\{1,...,i,...N\}$ is the set of active targets; $i$ denotes one target; $R_i^m$ and $P_i^m$ are the defender's reward and penalty, respectively, for an attack on this target; and $C_i^a$ and $C_i^m$ are the resources required to be expended by the attacker and the defender, respectively, to best protect target $i$. $A$ is the attacker, who commits to a strategy $\textbf{\textit{p}} = \{p_1,p_2,...,p_N\}$, where $p_i$ is the probability of an attack on target $i$. $D$ is the defender, who commits to a strategy $\textbf{\textit{q}} = \{q_1,q_2,...,q_N\}$, where $q_i$ is the probability of protecting target $i$. We take $\sum_{i \in T} {{q_i}\mathop C\nolimits_i^m \le \mathop M }$ to represent the constraint of the defender's available resources, where $q_iC_i^m$ represents the resources allocated to target $i$ and $M$ represents the maximum quantity of available resources.

\begin{table}[h]
\centering
\caption{\bf Parameter descriptions}
\begin{tabular}{|c|l|l}
\hline
\multicolumn{1}{|l|}{\bf Parameter } & \multicolumn{1}{|l|}{\bf Description} \\ 
\hline
$T$ & set of targets \\ 
$N$ & number of targets in $T$ \\ 
$A$ & attacker \\ 
$D$ & defender \\ 
$p_i$ & attack probability for target $i$ \\ 
$q_i$ & defense probability for target $i$ \\ 
$C_i^m$ & resources allocated to protect target $i$ \\ 
$M$ & maximum quantity of available resources\\ 
\hline
\end{tabular}
\label{tab1}
\end{table}

In this way, our problem is transformed into computing a reasonable defense probability distribution subject to the defender's resource constraints based on known parameters, including the resource constraints, the number of targets, the reward for protection, the penalty of protection, the cost of protection and the cost of attack for the set of targets.

\section{Model Formulation}\label{sec:model}
\begin{figure}[!h]
\centering
\includegraphics[scale = 0.6]{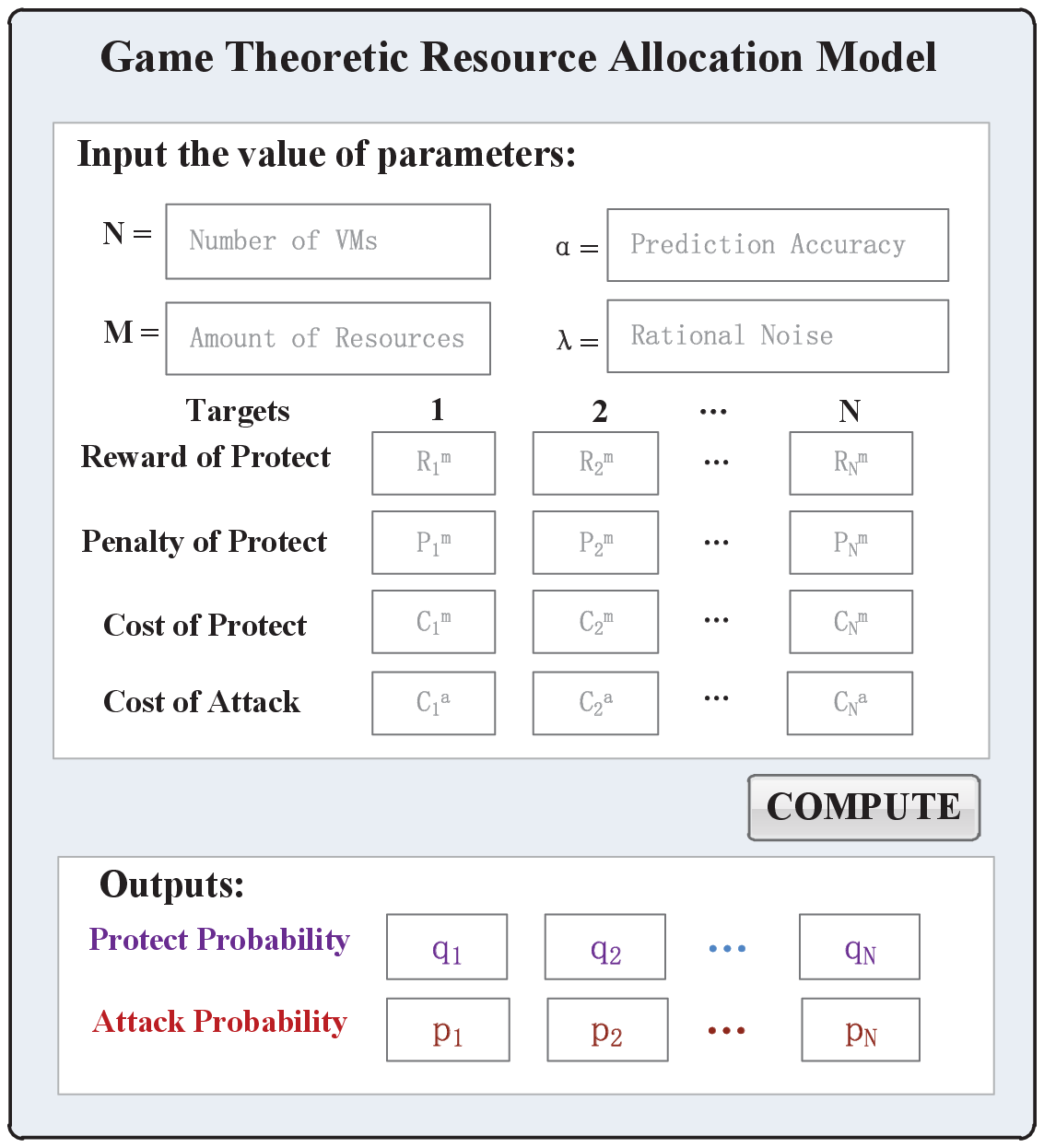}
\caption{\textbf{The game theoretic resource allocation interface.}\label{fig1}}
\end{figure}

To solve the given problem, we construct a Game Theoretic Resource Allocation (GTRA) model, as shown in Fig. \ref{fig1}. The input parameters are discussed in the above section. After the parameters are input, the proposed GTRA model computes the defender's possible defense probability distribution and the attacker's possible attack probability distribution.

For the computation process, a Stackelberg game is used to model the interaction between the defender and attacker. Then, the game payoff functions are built from the input parameters and are designed as strategic rules. Next, the QR model is used to simulate an attacker of the adversarial nature. In addition, an iterative genetic algorithm is utilized to obtain the equilibrium game strategy.

\subsection{Stackelberg Game}

Game theory \cite{11} is widely used to analyze problems in which all players who are in a conflict with a payoff attempt to win or to maximize their payoffs via changing their strategies based on the reactions of their adversaries. A Stackelberg game is a common game instance in which players select strategies sequentially: the leader moves first, and the follower responds accordingly.

In this paper, defender and attacker are the two rival roles. They are in conflict over the targets' security, and both attempt to maximize their own payoffs by allocating the fewest resources to the targets. Through game theoretical deduction, the defender first decides how to allocate resources to cover the targets; then, the attacker selects the targets to attack after observing the defender's strategy. The rivalry, the pursuit of the maximum payoffs and the sequence of actions make our problem fit perfectly into the framework of a Stackelberg game; thus, the GTRA model is built based on a Stackelberg game.

In a Stackelberg game, each player selects the action with the greatest payoff, which is defined as the player's return after taking the selected action. This payoff usually consists of reward, penalty and cost. In the proposed GTRA model, both the defender and the attacker can take two actions, so their payoffs for a target $i$ can be represented by a $2 \times2$ payoff matrix, as shown in Table \ref{tab2}. Clearly, there are four cases corresponding to the attacker's two actions (Attack or Not) and the defender's two actions (Protect or Not), which are represented by the four cells. Each cell contains two values separated by a comma: the first is the attacker's payoff, and the second is the defender's. In contrast to previous payoff matrices, we include action cost to measure the resource allocation metric directly.

\begin{table}[bhtp]
\footnotesize
\newcommand{\tabincell}[2]{\begin{tabular}{@{}#1@{}}#2\end{tabular}}
\caption{\textbf{Payoffs of the two players for target \textit{i}}}
\label{tab2}
\centering
~\\
\begin{tabular}{|c|c|c|c|}
\hline
~ & \tabincell{c}{\textbf{Protect $(q_i)$}}&\tabincell{c}{\textbf{ Not Protect $(1-q_i)$ }} \\
\hline
\tabincell{c} {\textbf{Attack}\\  $(p_i)$} & \tabincell{c}{$-\alpha P_i^a+(1-\alpha) R_i^a-C_i^a$, \\ $\alpha P_i^a-(1-\alpha) R_i^a-C_i^m$}& \tabincell{c}{$R_i^a-C_i^a$,\\ $-R_i^a$}\\
\hline
\tabincell{c} {\textbf{Not Attack} \\ $(1-p_i)$} & {0 , $-C_i^m$ }&{0 , 0}\\
\hline
\end{tabular}
\end{table}

{\bf Case 1:} $\{$Attack, Protect$\}$. The attacker launches an attack on target $i$, and the defender protects it simultaneously. In this case, the attacker's benefit is $-\alpha P_i^a + (1 - \alpha )R_i^a$, and the defender's benefit is $\alpha P_i^a + (1 - \alpha )R_i^a$, where $\alpha$ is the accuracy of attack prediction, $P_i^a$ is the attack penalty, and $R_i^a$ is the attack reward.

{\bf Case 2:} $\{$Attack, Not Protect$\}$. The attacker launches an attack on target $i$, and the defender does not protect it. In this case, the attacker will not be punished, and its payoff is the difference between $R_i^a$ and the cost. The defender's payoff is $-R_i^a$ alone, without a cost, because no protection is attempted. 

{\bf Case 3:} $\{$Not Attack, Protect$\}$. The attacker does not launch an attack on target $i$, but the defender protects it. In this case, the attacker's payoff is zero due to the absence of an attack, and the defender's payoff is the negative value corresponding to the cost of the consumed security resources, $-C_i^m$, with no benefit.

{\bf Case 4:} $\{$Not Attack, Not Protect$\}$. The attacker does not launch an attack on target $i$, and the defender does not protect it. In this case, each player's payoff is zero because neither performs an action.

To distinguish different targets, one work \cite{3} considered targets with different noncorrelated security assets. Motivated by that study, we label targets with different security assets in the form of distinct rewards, penalties and action costs for the defender and attacker. A player's total payoff is the combination of the four separate cases. In combination with the attack probability, defense probability and payoff items shown in Table \ref{tab2}, the total payoff functions of the defender and the attacker are given in \eqref{equ:1} and \eqref{equ:2}, respectively. These two utility functions are different from the utility functions used in many previous studies because the players' action costs are directly included in our utility functions.

\begin{align}
\begin{split}
&\scalebox{0.9}{$U_M = \sum\limits_{i \in {\rm T}}{{{p_i}{q_i}[\alpha P_i^a-(1-\alpha )R_i^a-C_i^m]}-{p_i}(1-{q_i})R_i^a}$}
\\
&\scalebox{0.9}{$-(1-{p_i}){q_i}C_i^m = \sum\limits_{i \in {\rm T}} {{q_i}[\alpha {p_i}(P_i^a + R_i^a) - C_i^m] - {p_i}R_i^a}$}
\end{split}
\label{equ:1}
\end{align}

\begin{align}
\begin{split}
&\scalebox{0.9}{$ {U_A = \sum\limits_{i \in {\rm T}} {{p_i}} {q_i}[ - \alpha P_i^a + (1 - \alpha )R_i^a - C_i^a]} + {p_i}(1 - {q_i})*$}
\\
&\scalebox{0.9}{$ (R_i^a - 
C_i^a)
= \sum\limits_{i \in {\rm T}} {{p_i}} [ - \alpha {q_i}(P_i^a + R_i^a) + (R_i^a - C_i^a)]$}
\end{split}
\label{equ:2}
\end{align}

For both the defender and the attacker, the objective of each player is to maximize that player's own payoff by designing an optimal strategy. When both players achieve their maximum payoffs, the corresponding solution to the problem is called the Nash equilibrium \cite{11}. 

{\bf Definition} Consider a game $G=\{s_1,..,s_n;u_1,...,u_n\}$ with $n$ players. If, for a strategy profile $\{s_1^*,...,s_n^*\}$, the strategy $s_i^*$ for every player $i$ is either the optimal strategy for that player or a strategy that is no worse than any of the other $(n-1)$ strategies, then that strategy profile is called a Nash equilibrium (NE) strategy profile.

The NE of a Stackelberg game can be derived by applying backward induction \cite{2}, which involves  reasoning from the end of a situation to determine the sequence of optimal strategies. In this context, we deduce the defender's protection strategy in a forward manner from the attacker's situation in each round, as follows.

Follower: Attacker side. The attacker observes the defender's strategy and designs a greedy strategy to maximize its payoff. Formally, for any given $\textbf{\textit{q}} \in S_M$, the attacker's task is to solve the optimization problem in \eqref{equ:3}. 

\begin{eqnarray}
\begin{split}
& \textbf{\textit{p}}(\textbf{\textit{q}}) = arg max {U_A}(\textbf{\textit{p}},\textbf{\textit{q}}(\textbf{\textit{p}}))  
\\
& Subject\;to \qquad {\rm{0}} \le {p_i} \le {\rm{1}} 
\end{split}
\label{equ:3}
\end{eqnarray}

Leader: Defender side. The defender knows that the attacker will respond greedily. Therefore, the defender designs a protection strategy based on the attacker's potentially best response. Formally, the defender needs to solve the optimization problem in \eqref{equ:4}. The first constraint suggests that the total quantity of resources available to the defender is no more than $M$. 

\begin{eqnarray}
\begin{split}
 \textbf{\textit{q}}(\textbf{\textit{p}}) = arg max {U_M}(\textbf{\textit{p}}(\textbf{\textit{q}}),\textbf{\textit{q}})
\\
 Subject\;to  \qquad \sum\limits_{i \in T} {{q_i}C_i^m} \le {M}{\rm{ }} 
 \\
  and \; \; \qquad  \qquad  {\rm{0}} \le {q_i} \le 1 
\end{split}
\label{equ:4}
\end{eqnarray}

We derive the NE from the above two sequential steps. We derive $\textbf{\textit{q}}^\textbf{*}$ by solving \eqref{equ:4}; then, $\textit{\textbf{p}}^*$ is derived as $\textbf{\textit{p}}(\textbf{\textit{q}}^*)$ by solving \eqref{equ:3}. Finally, the strategy combination $(\textit{\textbf{p}}^*,\textit{\textbf{q}}^*)$ is the equilibrium game strategy for the Stackelberg game, and the final resources allocated to target $i$ will be $q^* * C_i^m$.

\subsection{Quantal Response}
The previous analysis is performed under the assumption that the attacker is perfectly rational and develops its strategy with complete knowledge of the defender's strategy. However, in the real world, the attacker will not always be perfectly rational since the attacker cannot always know the defender's strategy. Consequently, the defender is unsure whether the attacker will operate according to the predictive strategy $\textbf{\textit{p}(\textit{q})}$. If the attacker is not perfectly rational and chooses a strategy that deviates slightly from the rational strategy, the defender's payoff may decrease. Clearly, the defender is unwilling to accept a lower payoff while doing nothing.

To simulate the bounded rational adversary, many behavior models have been proposed, including quantal response (QR), SUQR, prospect theory (PT) and so on, which are all commonly used. The defender's response to them in the game theoretic model has been done in our previous work \cite{liu2017response} and we found the defender's response to the QR model is the most careful where the defense probability is relatively bigger than the other two bounded rational models. Hence, the QR model is introduced into the GTRA model to simulate the attacker's adversarial nature and to improve the proposed model.

When the QR model is applied, the noise in a bounded rational attacker's strategy is controlled by $\lambda$. $\lambda=0$ represents a uniform random probability distribution over the attacker's possible strategies, while $\lambda \xrightarrow{}\infty$ represents a perfectly rational attacker. Thus, the attacker's probability of attacking target $i$ is changed to \eqref{equ:5}.

\begin{eqnarray}
{p_i} = \frac{{{e^{\lambda {U_A}({q_i})}}}}{{\sum\nolimits_{j = 1}^n {{e^{\lambda {U_A}({q_j})}}} }}\
\label{equ:5}
\end{eqnarray}

Furthermore, the defender's utility function becomes \eqref{equ:8}.

\begin{align}
\begin{split}
&\scalebox{0.9}{$U_M = \sum\limits_{i \in T} {{q_i}}  * [\alpha {p_i}(\mathop P\nolimits_i^a  + \mathop R\nolimits_i^a ) - \mathop C\nolimits_i^m ] - {p_i}\mathop R\nolimits_i^a$}
\\
&\scalebox{1}{$= \sum\limits_{i \in T} {[\frac{{[\alpha {q_i}(\mathop P\nolimits_i^a  + \mathop R\nolimits_i^a ) - \mathop R\nolimits_i^a ] * \mathop e\nolimits^{ - \lambda [\alpha {q_i}(\mathop P\nolimits_i^a  + \mathop R\nolimits_i^a ) - (\mathop R\nolimits_i^a  - \mathop C\nolimits_i^a )]} }}{{\sum\nolimits_{j = 1}^n {\mathop e\nolimits^{ - \lambda [\alpha{q_j}(\mathop P\nolimits_j^a  + \mathop R\nolimits_j^a ) - (\mathop R\nolimits_j^a  - \mathop C\nolimits_j^a )]} } }} - {q_i}\mathop C\nolimits_i^m ]} $}
\end{split}
\label{equ:8}
\end{align}

In summary, the proposed GTRA model is used to solve the defender's problem of how to allocate $M$ units of resources to maximize the defender's utility function, as illustrated in \eqref{equ:9}.
\begin{eqnarray}
 _\textbf{\;\;\textit{q}}^{\max }\;U_M  \; \qquad  \qquad {\rm{ s}}{\rm{.t}}{\rm{.}}\left\{ \begin{array}{l}
 \sum\limits_{i \in T} {{q_i} C_i^m \le M } \\
0 \le {q_i} \le 1,{\rm{ }}\forall i
\end{array} \right.  
\label{equ:9}
\end{eqnarray}

\subsection{Algorithm}

Since the defender's objective utility function expressed in \eqref{equ:9} corresponds to a nonlinear constraint problem, the optimal solution is extremely difficult to find. As a classic algorithm for searching for an approximately optimal solution \cite{NFV}, the genetic algorithm (GA) provides an alternative approach. GA is a stochastic global search and optimization method that mimics natural biological evolution. However, the typical GA attempts to find a globally near-optimal solution instead of a globally optimal one. Therefore, in this paper, we utilize Algorithm \ref{algorithm1} to compute the defender's equilibrium strategy in the proposed GTRA model.

\begin{algorithm}[!h]
\footnotesize
\caption{Iterative Genetic Algorithm (IGA)}
\label{algorithm1}
\begin{algorithmic}[1]
\STATE \textbf{Initialization:}
\\number of targets $\rightarrow N$;
\\number of iterations $\rightarrow times = 10$;
\\resource constraint $\rightarrow M$;
\\$Ud^* \leftarrow -\infty$
\STATE \textbf{Iteration:}
\WHILE{$i < times $}
\STATE $(q_i,Ud_i) \leftarrow GA(MultiObj, N, M)$
\IF{$Ud_i > Ud^*$}
\STATE $Ud^* = Ud_i$;
\STATE $q^* = q_i$;
\ENDIF
\ENDWHILE
\RETURN $(q^*,Ud^*)$
\end{algorithmic}
\end{algorithm}

In addition to the parameters of the utility function discussed in the previous section, the number of targets, the number of iterations and the resource constraint are initialized before the iteration process. In each iteration, we find the locally optimal strategy $q_i$ and the corresponding utility $Ud_i$ using the $GA()$ function in MATLAB. Then, we record the current maximum after each iteration. When the iteration number $i$ reaches the given maximum $times$, the globally optimal strategy $q_i^*$ and the corresponding utility $Ud_i^*$ are obtained. In general, the probability of reaching the global optimum increases as the number of iterations $times$ increases.

To better understand the equilibrium game strategy, we illustrate the evolutionary behavior of the defender by adopting the phase plane of replicator dynamics \cite{replicator}. First, tersely describe the payoff of the defender and attacker in every case, Table \ref{tab2} is changed into Table \ref{tab3} where $a = -\alpha P_i^a+(1-\alpha) R_i^a-C_i^a$, $b = \alpha P_i^a-(1-\alpha) R_i^a-C_i^m$, $c = R_i^a-C_i^a$, $d = -R_i^a$ and $f = -C_i^m$. Then, the replicator dynamics equations of the attacker and the defender are expressed as \eqref{dr}. $\bar{U}_A$ and $\bar{U}_M$ represent the average payoffs. The evolutionary equilibrium can then be obtained by solving the following equations $\dot{p}=0$ and $\dot{q}=0$.

\begin{table}[!h]
\centering
\caption{\bf Simplified payoffs for target $i$}
\begin{tabular}{|c|c|c|}
\hline
\multicolumn{1}{|l|}{\bf ~ } & \multicolumn{1}{|l|}{\bf Protect $(q)$}& \multicolumn{1}{|l|}{\bf Not Protect $(1-q)$} \\ \hline
Attack ($p$) & $a,b$ & $c,d$ \\
\hline
 Not Attack ($1-p$) & $0,f$ & $0,0$ \\
\hline
\end{tabular}
\label{tab3}
\end{table}

\begin{eqnarray}
\setlength{\abovedisplayskip}{0pt} 
\begin{split}
\label{dr}
& \dot{p}=p(U_A - \bar{U}_A)=p(1-p)[qa+(1-q)c]
\\
& \dot{q}=q(U_M - \bar{U}_M)=q(1-q)[p(b-d)+(1-p)f]
\end{split}
\end{eqnarray}

\begin{figure}[!b]
\centering
\includegraphics[scale =  0.41]{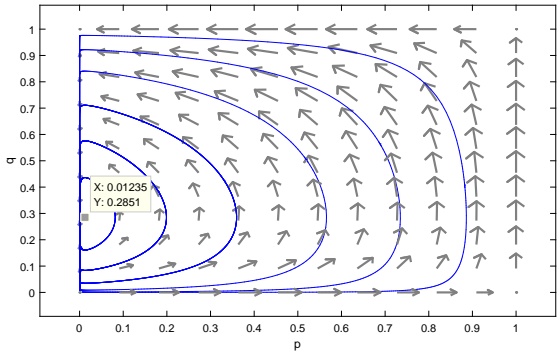}
\caption{\textbf{Evolution process of the defender's behavior.}\label{fig3}}
\end{figure}

Fig. \ref{fig3} presents the evolutionary equilibrium of the defender's strategy, which can be seen as the process of adapting the initial strategy to the NE strategy. The smallest circle around the NE point $(0.012,0.2851)$ is the entire feasible region of the solution.

\section{Numerical Study}\label{sec:experiment}
Since the focus of this paper is to explore the impact of different parameter configurations, we perform the numerical analysis directly to validate the proposed method. We first compare the proposed utility function with the utility function that does not consider the action cost. Then, we compare the NE strategy that is computed based on the proposed utility function with four other resource allocation strategies. In each group of experiments, 100 game instances under the same conditions are considered, and the average value is taken as the result. In each game instance, the number of iterations in Algorithm \ref{algorithm1} is set to 10 \footnote{We conducted an experiment with 100 iterations and found that the maximum value was usually found within the first ten iterations.}, and the maximum value is taken. 

\subsection{Comparison of Utility Functions}
To assess the impact of the action cost on the strategy, we compare the utility functions with and without action cost, respectively. Specifically, to explore the influence of the relationship between the two players' action costs on each player's strategy, we design three groups of experiments, as shown  in Table \ref{tab:Ufuncs}.

\begin{table}[!h]
\centering
\caption{\bf Four utility function scenarios}
\begin{tabular}{|c|c|c|c|}
\hline
{No. } &{Relation}&{$C^m$}& {$C^a$} \\ \hline
1 & $C_i^m > C_i^a$ & $\gamma< C_i^m <2*\gamma$ & $ 0< C_i^a <\gamma$\\
\hline
2 & $C_i^m < C_i^a$ & $ 0< C_i^m <\gamma$ & $\gamma< C_i^a <2*\gamma$ \\
\hline
3 & $C_i^m = C_i^a$ & $ 0< C_i^m <\gamma$ & $C_i^a = C_i^m$ \\
\hline
4 & $NoCost$ & $\gamma =0$ & $\gamma =0$ \\
\hline
\end{tabular}
\label{tab:Ufuncs}
\end{table}

\noindent The first scenario corresponds to a utility function in which the cost of defense is greater than the cost of attack. On the contrary, the second scenario corresponds to a utility function in which the cost of attack is greater than the cost of defense. The third scenario corresponds to a utility function in which the costs of attack and the cost of defense are equal and are greater than 0. The fourth scenario corresponds to the utility function used in previous works \cite{alert,securityresource,lim3,lim6,reward,16}, in which the action costs $C_i^a$ and $C_i^m$ are $0$ and the resource consumption is simply the sum of the defense probabilities.

\begin{figure}[!t]
\centering
\subfigure[Defender's utility comparison (ratio = 0.1)]{\includegraphics[scale=0.25]{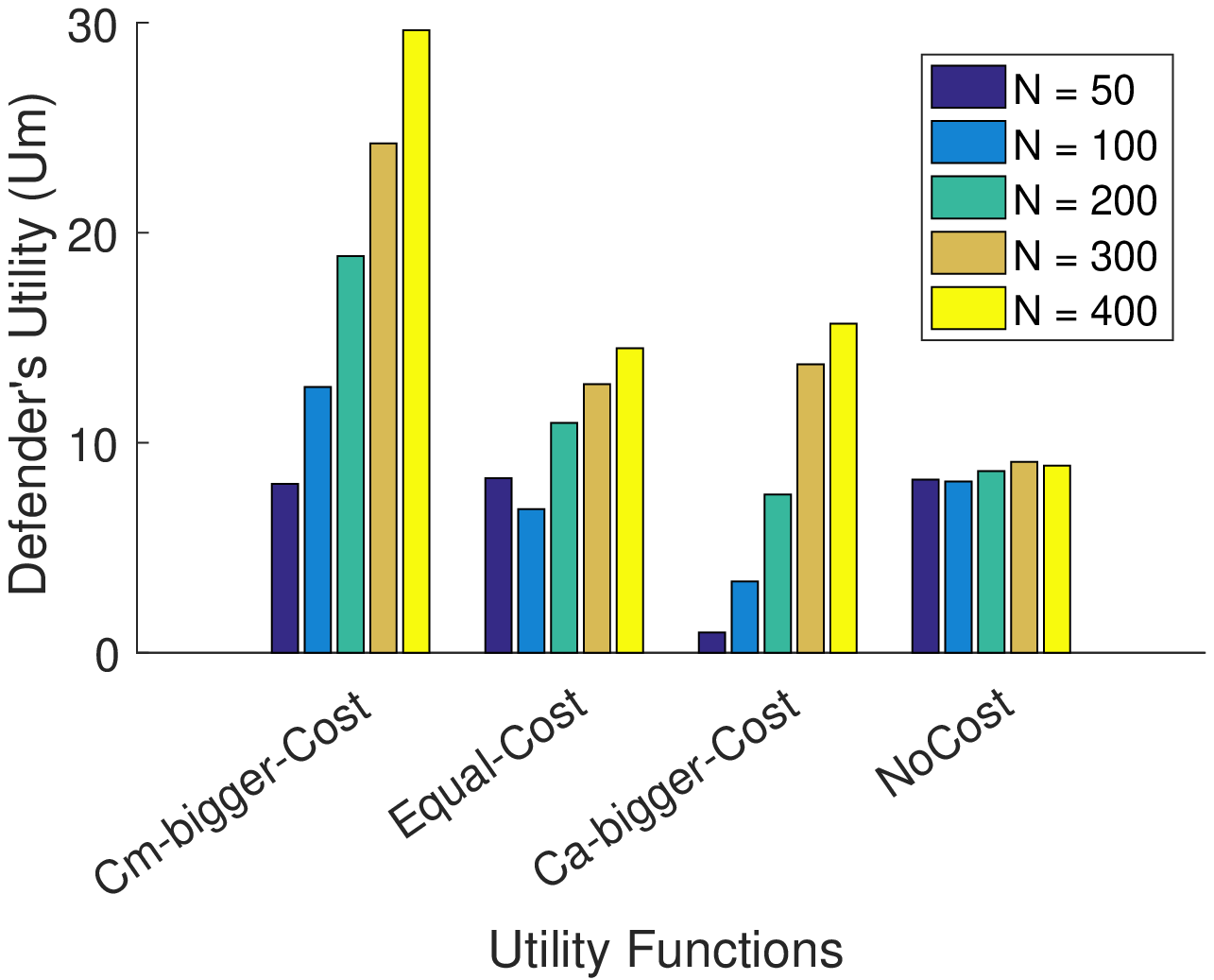}
\label{subfig:Um_n}}
\hfil
\subfigure[Defender's utility comparison (N = 200)]{\includegraphics[scale=0.25]
{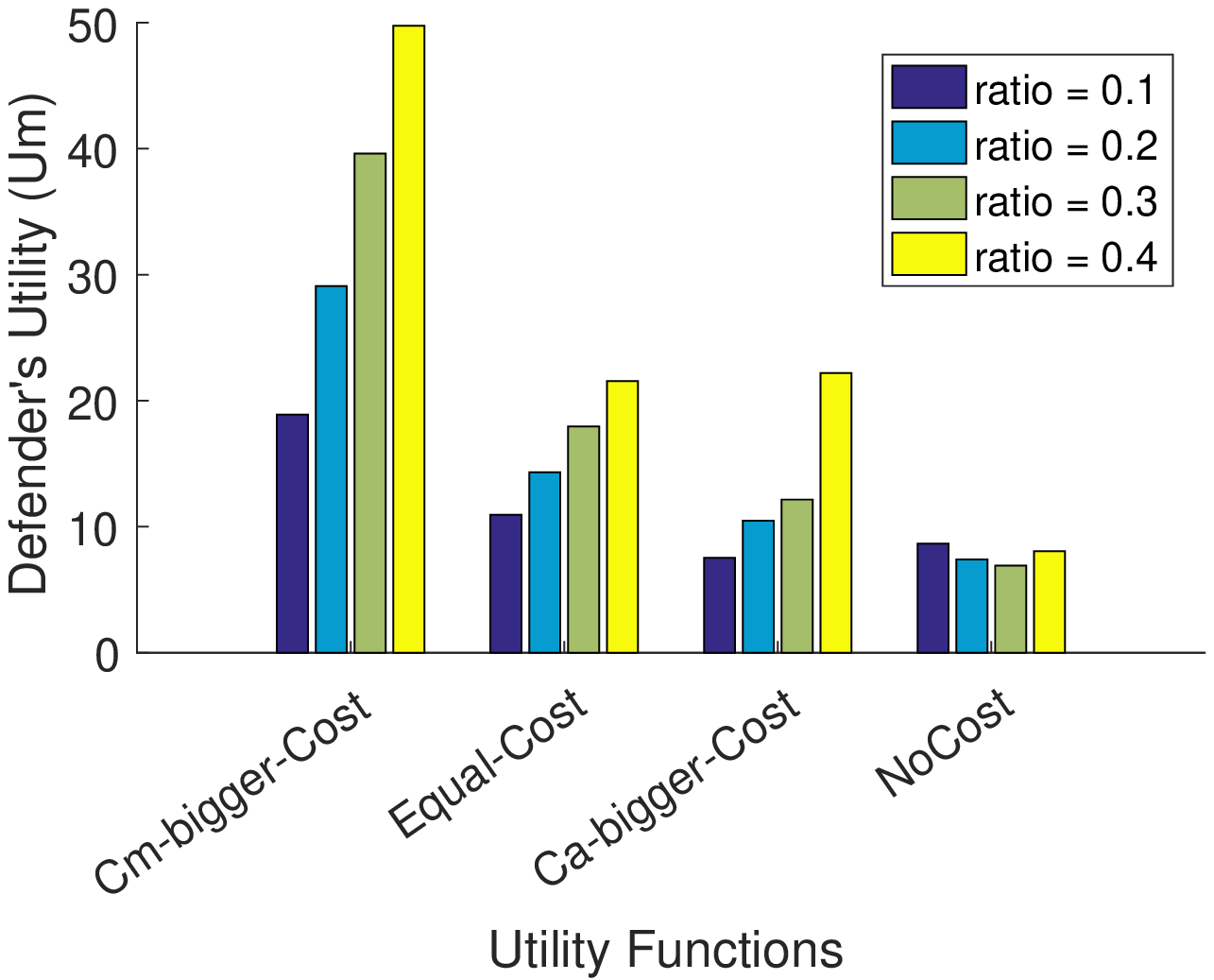}
\label{subfig:Um_r}}
\hfil
\subfigure[Resource comparison (ratio = 0.1)]{\includegraphics[scale=0.25]{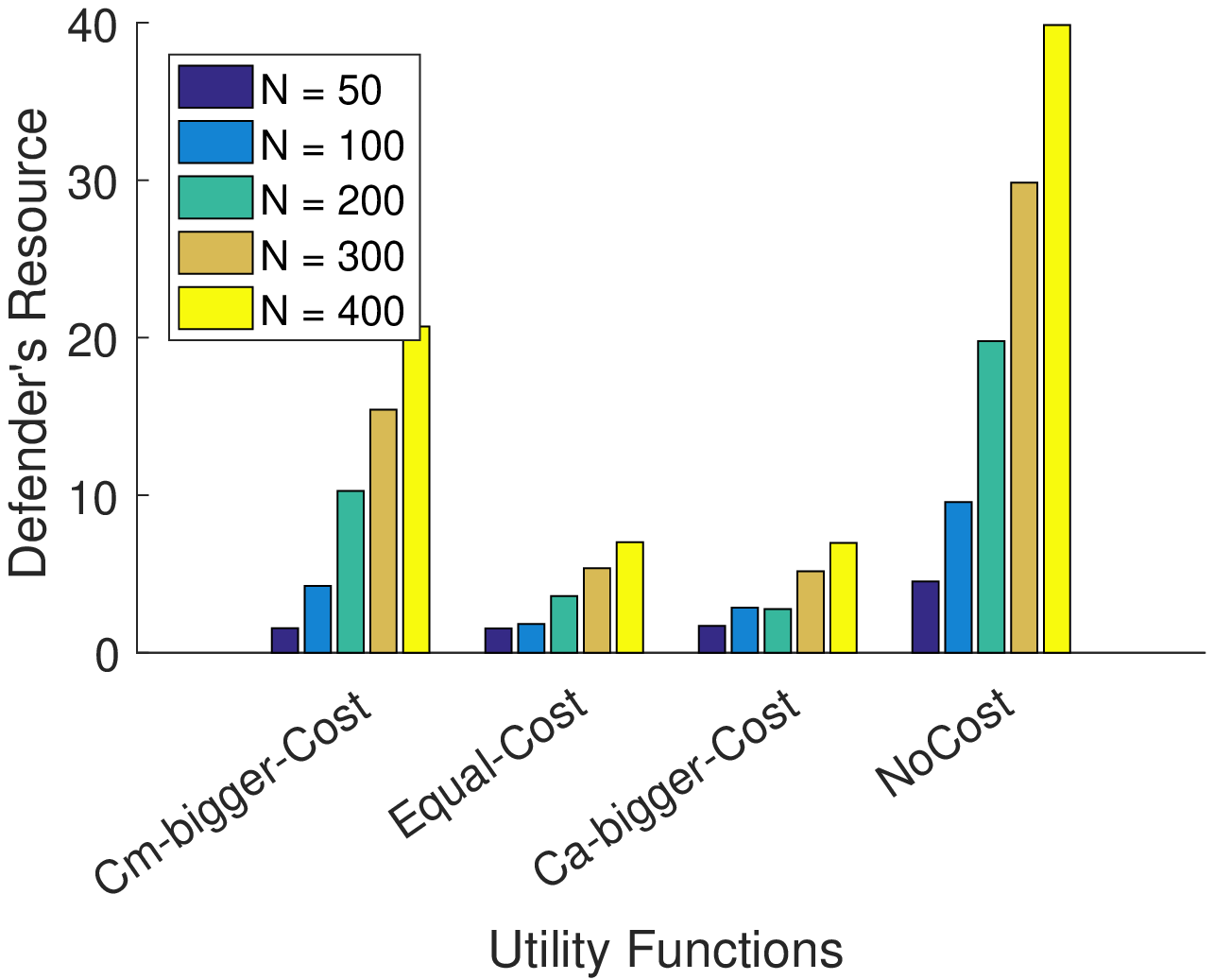}
\label{subfig:Re_n}}
\hfil
\subfigure[Resource comparison (N = 200)]{\includegraphics[scale=0.25]
{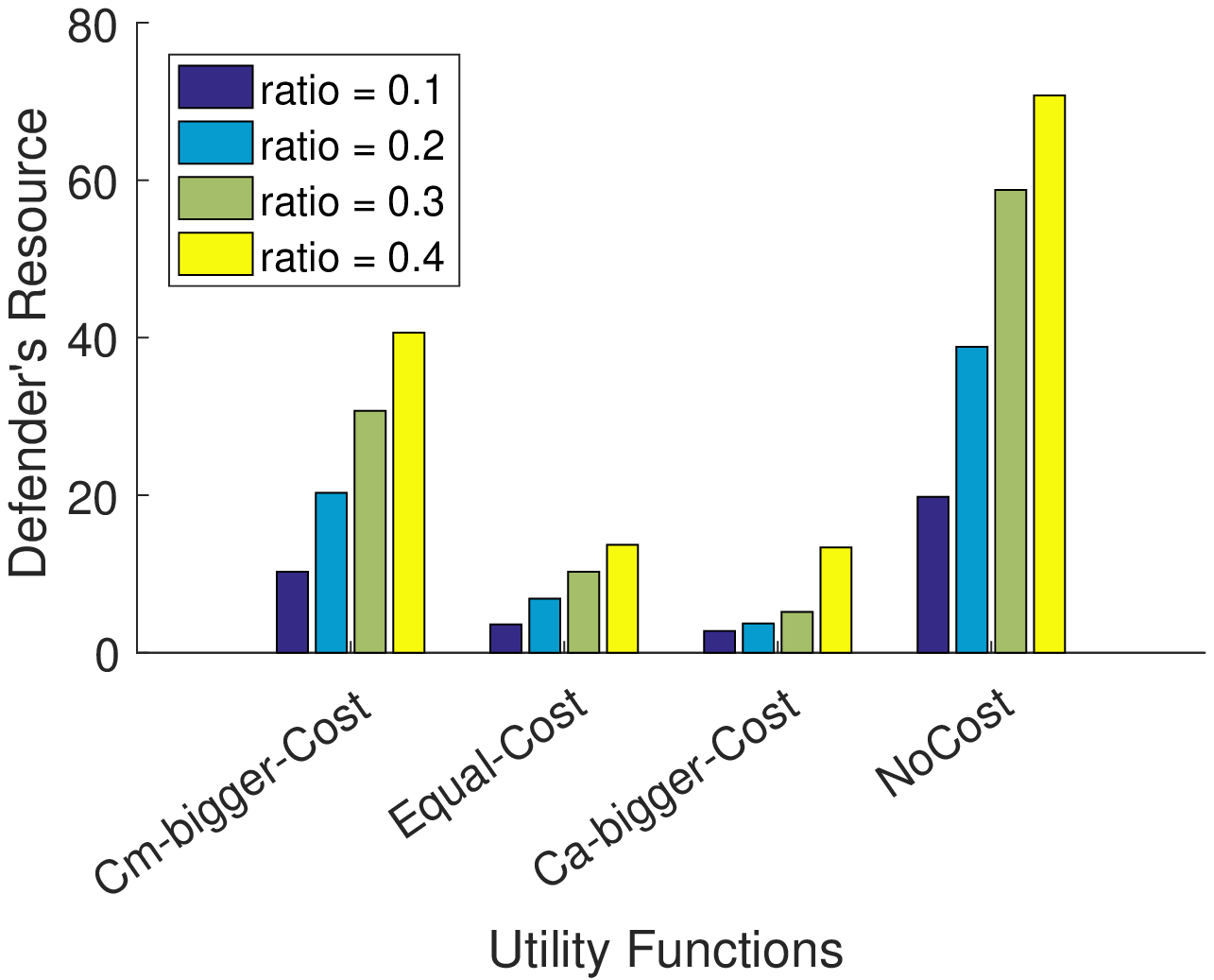}
\label{subfig:Re_r}}
\hfil
\subfigure[Effectiveness comparison (ratio = 0.1)]{\includegraphics[scale=0.25]{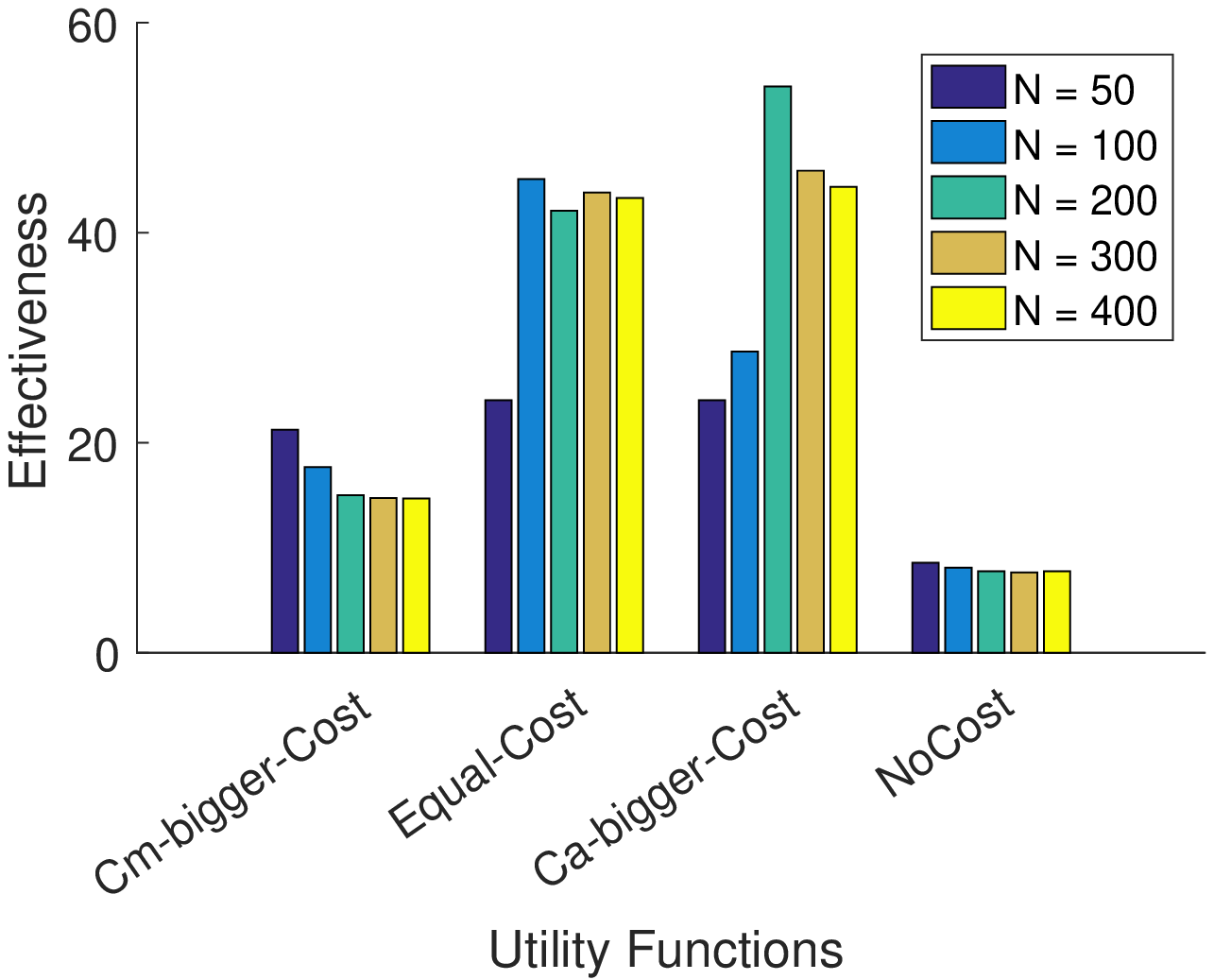}
\label{subfig:Effect_n}}
\hfil
\subfigure[Effectiveness comparison (N = 200)]{\includegraphics[scale=0.25]
{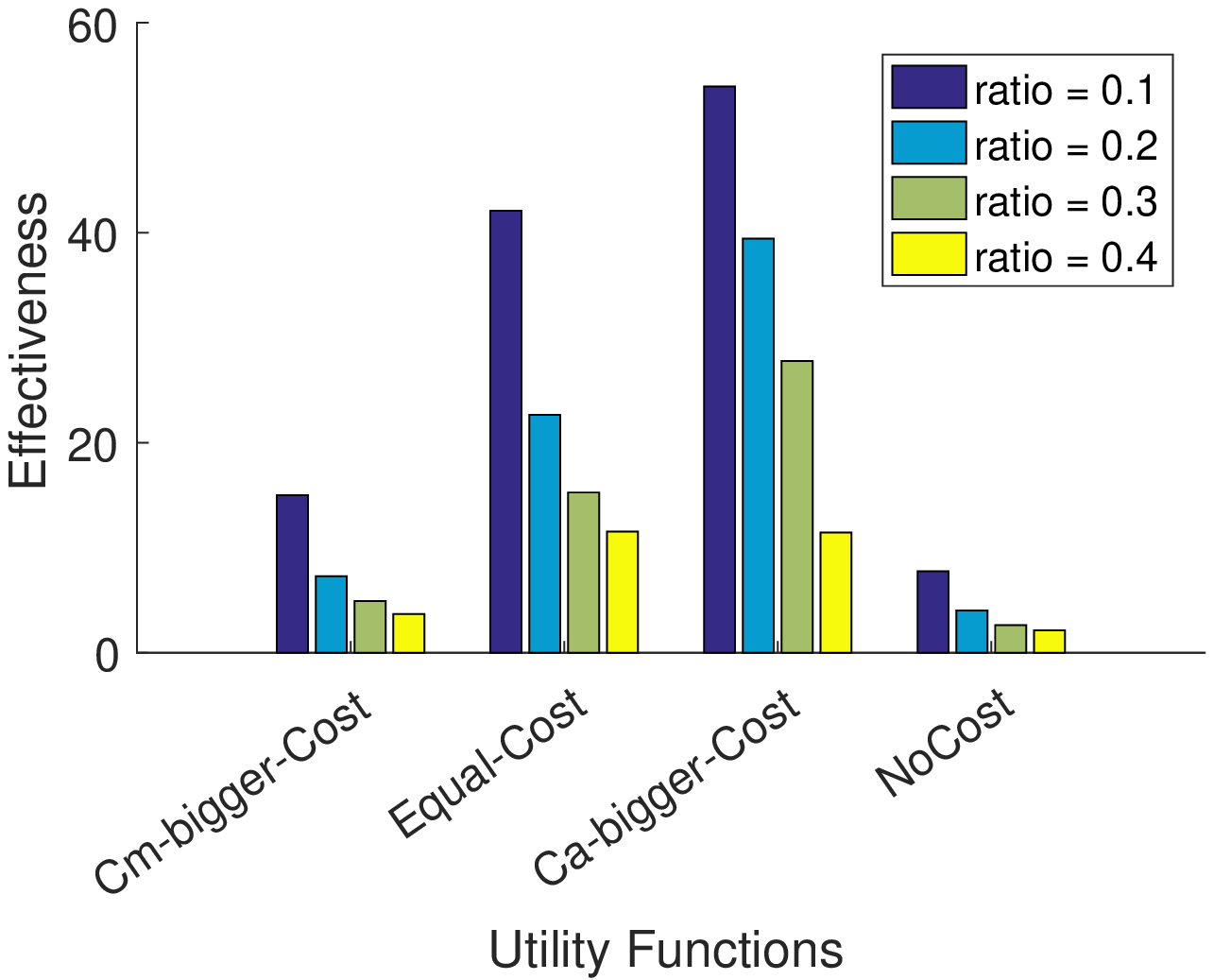}
\label{subfig:Effect_r}}
\caption{\textbf{Solution quality comparison of different utility functions.}}
\label{fig:Solution}
\end{figure}

We measure the solution quality in each scenario in terms of the defender's average utility and the average effectiveness over all 100 game instances, where the \textit{effectiveness} is defined as the average number of protected targets per resource, as shown in \eqref{equ:eff}. The growth rate, defined in \eqref{equ:effGrowth}, is used to measure the difference in solution quality between different utility function scenarios, where $a$ denotes the solution quality for a utility function without action cost and $b$ denotes the solution quality for a utility function that includes the action cost. 

The parameters used in this paper are the same as those used in the previous study \cite{reward}, where reward $R_i^a$ is chosen randomly from a uniform distribution from 1 to 10,  penalty $P_i^a$ is chosen randomly from a uniform distribution from -10 to -1 and the resource constraint is proportional to the number of targets, $M = \gamma * N$. We assume that the total available resources, including the total cost, are insufficient to protect all targets. Therefore, the value of parameter $ \gamma$ is set to less than 1. 

\begin{eqnarray}
effectiveness = \frac{number\_of\_covered\_ targets}{consumed\_ resources}
\label{equ:eff}
\end{eqnarray}

\begin{eqnarray}
Growth\_Rate = \frac{a - b}{b}
\label{equ:effGrowth}
\end{eqnarray}

Fig. \ref{fig:Solution} shows the solution quality results for the various utility function scenarios introduced in Table \ref{tab:Ufuncs} with different parameter configurations. The defender's average utility, and the defender's resource consumption along with the effectiveness are displayed on the y-axes. On the x-axes, Figs. \ref{subfig:Um_n}, \ref{subfig:Re_n} and \ref{subfig:Effect_n} show the results of varying the number of targets ($N$) while keeping the ratio ($\gamma$) of resources ($M$) to $N$ fixed to $0.1$. Figs. \ref{subfig:Um_r}, \ref{subfig:Re_r} and \ref{subfig:Effect_r} show the results of varying the ratio of resources to targets while keeping the number of targets fixed at $200$. The corresponding solution qualities in the various utility function scenarios are presented as groups of bars.

Figs. \ref{subfig:Um_n}, \ref{subfig:Re_n} and \ref{subfig:Effect_n} show the following. (1) The defender's utility ($U_m$) increases as the number of targets ($N$) increases. The utility is larger in the first scenario than other scenarios under the same conditions, and it is nearly stable in the fourth scenario, regardless of $N$, which indicates that the greater cost of defense has a better effect on obtaining payoff under the same conditions. (2) The defender's resource consumption increases as the number of targets ($N$) increases. The resource consumption is larger in the first scenario than in the second and third scenarios, and it is strongly proportional to $N$ in the fourth scenario. It illustrates the cumulative impact of action costs on a massive number of targets. (3) The effectiveness does not vary regularly with the number of targets ($N$); it varies inversely with the resource consumption in the first scenario, in which the action cost is considered in the utility function, while a nearly constant effectiveness is maintained in the fourth scenario. It suggests that when expending the same number of resources, the number of protected targets of the fourth scenario where the action costs are not considered in the utility function is the least.

Figs. \ref{subfig:Um_r}, \ref{subfig:Re_r} and \ref{subfig:Effect_r} show the following. (1) The defender's utility ($U_m$) increases as the ratio of resources to the number of targets increases. The defender's utility is larger than the second and third scenarios under the same conditions when $C_i^m > C_i^a$, and it is nearly stable when there is no action cost in the utility function, regardless of the resource-to-target ratio. It reveals that the utility functions including action cost provide more utility than those without action cost. Moreover, the number of resources has a positive effect on the defender's utility. (2) The defender's resource consumption increases as the resource-to-target ratio increases; it is larger in the first scenario than in the second and third scenarios and it is strongly proportional to the resource-to-target ratio in the fourth scenario. It also shows the cumulative impact of action costs on a massive number of targets. (3) The effectiveness decreases with an increasing resource-to-target ratio. The effectiveness varies inversely with resource consumption in all four scenarios, and it is smaller in the first scenario than in the second and third scenarios. It suggests that the effectiveness decreases with the increasing number of resources under the same conditions, and the effectiveness is the least when the utility does not include the action cost.

\begin{figure}[!t]
\centering
\subfigure[Defender's utility comparison (ratio = 0.1)]{\includegraphics[scale=0.25]{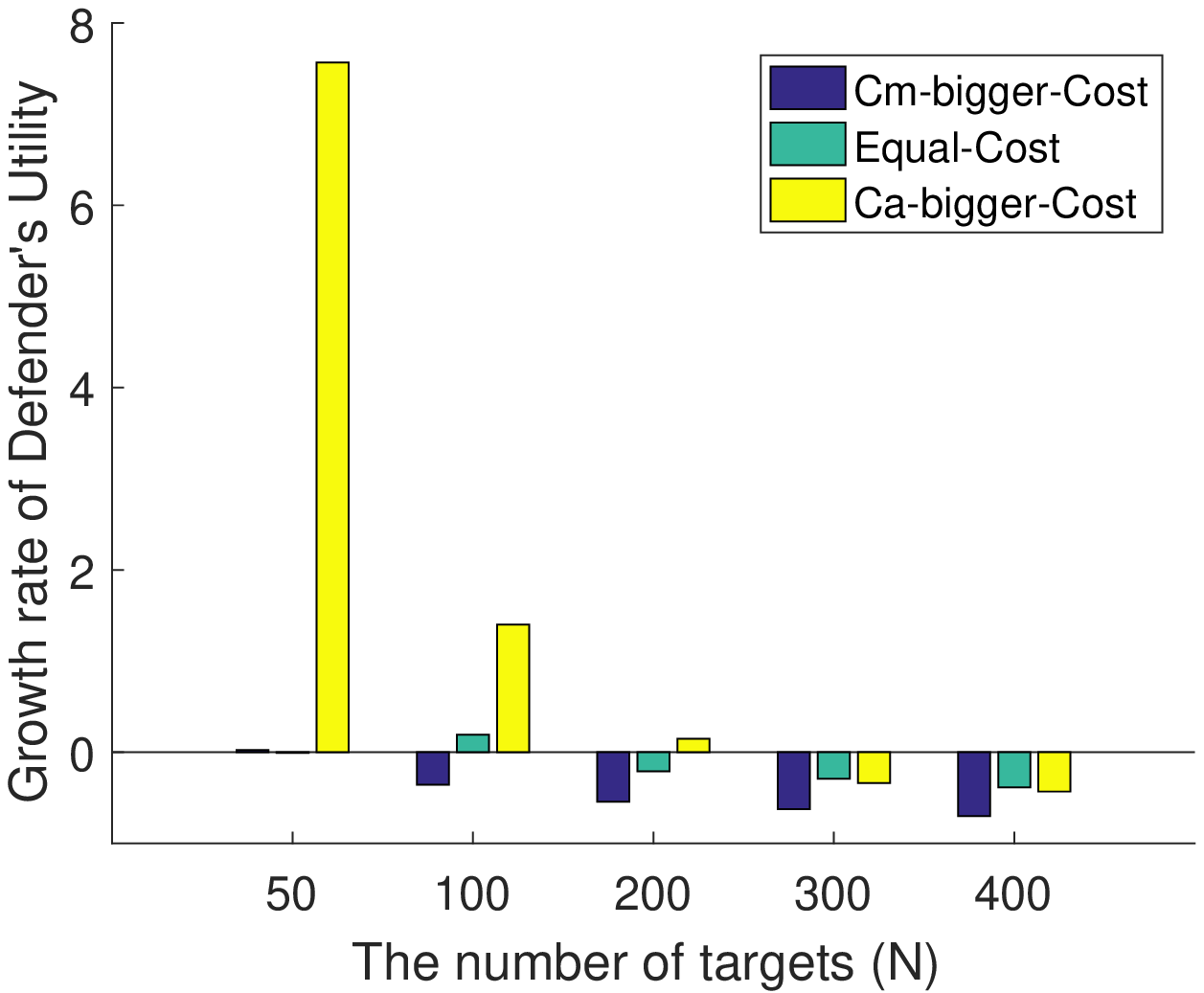}
\label{subfig:Um_nG}}
\hfil
\subfigure[Defender's utility comparison (N = 200)]{\includegraphics[scale=0.25]
{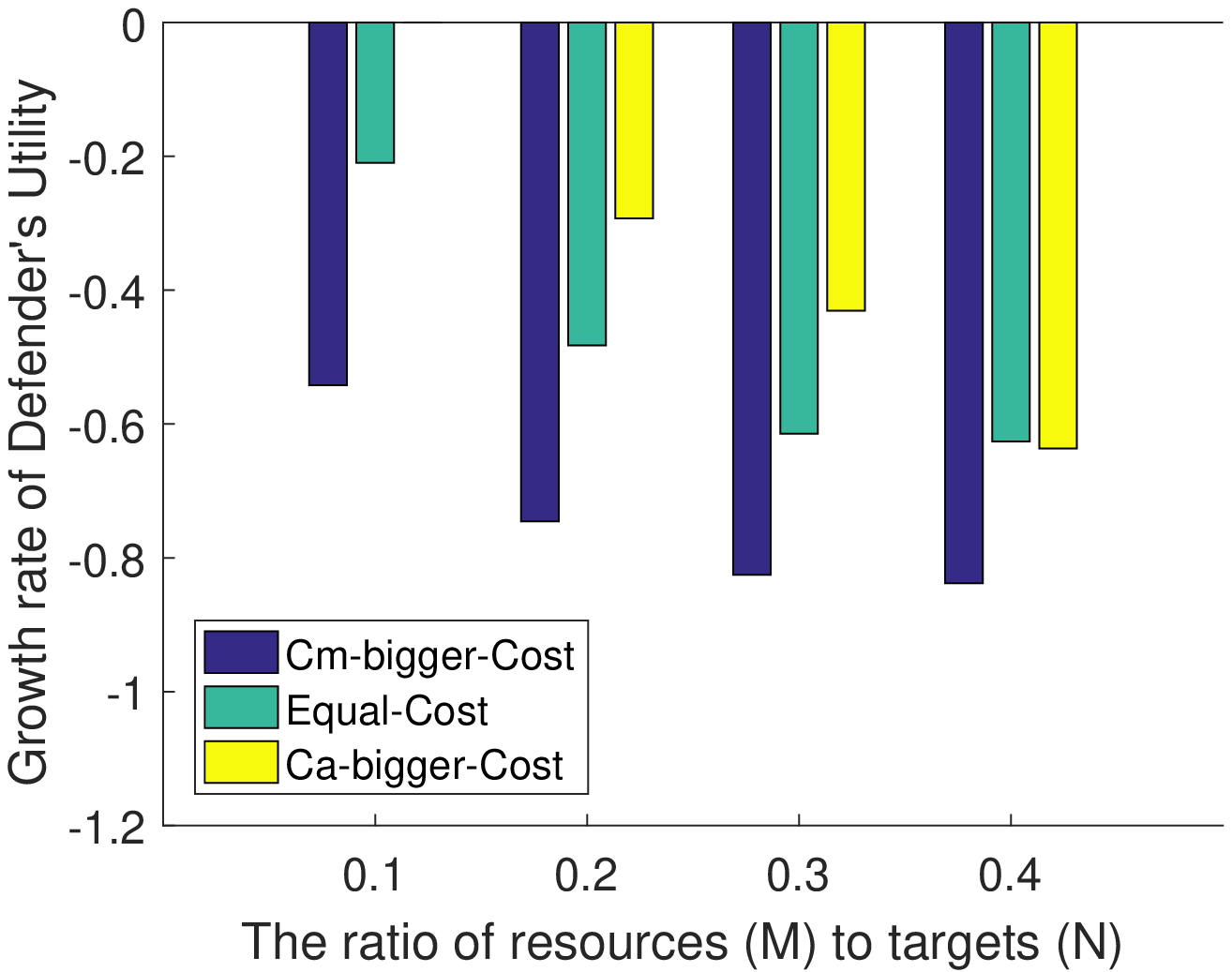}
\label{subfig:Um_rG}}
\hfil
\subfigure[Resource comparison (ratio = 0.1)]{\includegraphics[scale=0.25]{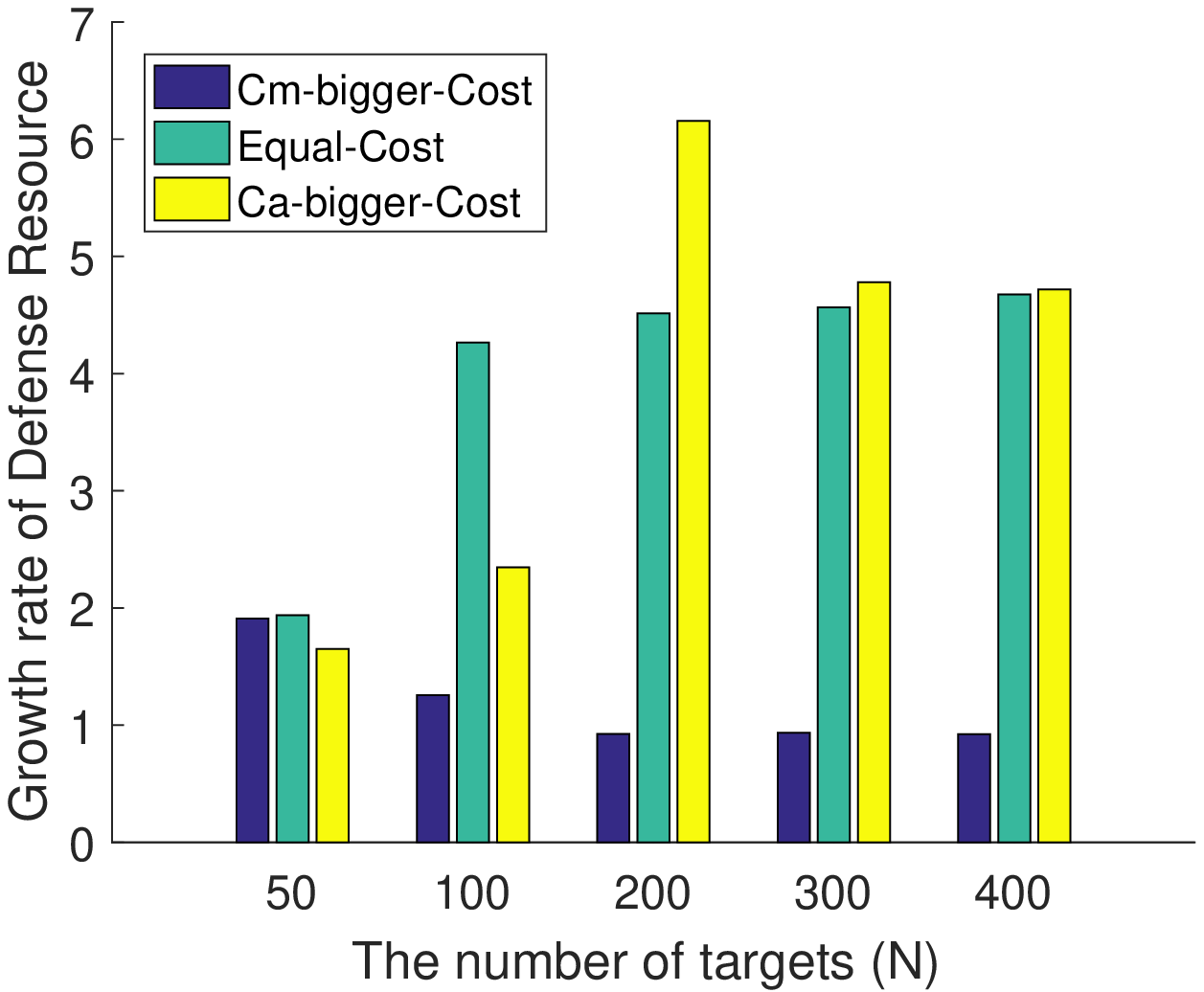}
\label{subfig:Re_nG}}
\hfil
\subfigure[Resource comparison (N = 200)]{\includegraphics[scale=0.25]
{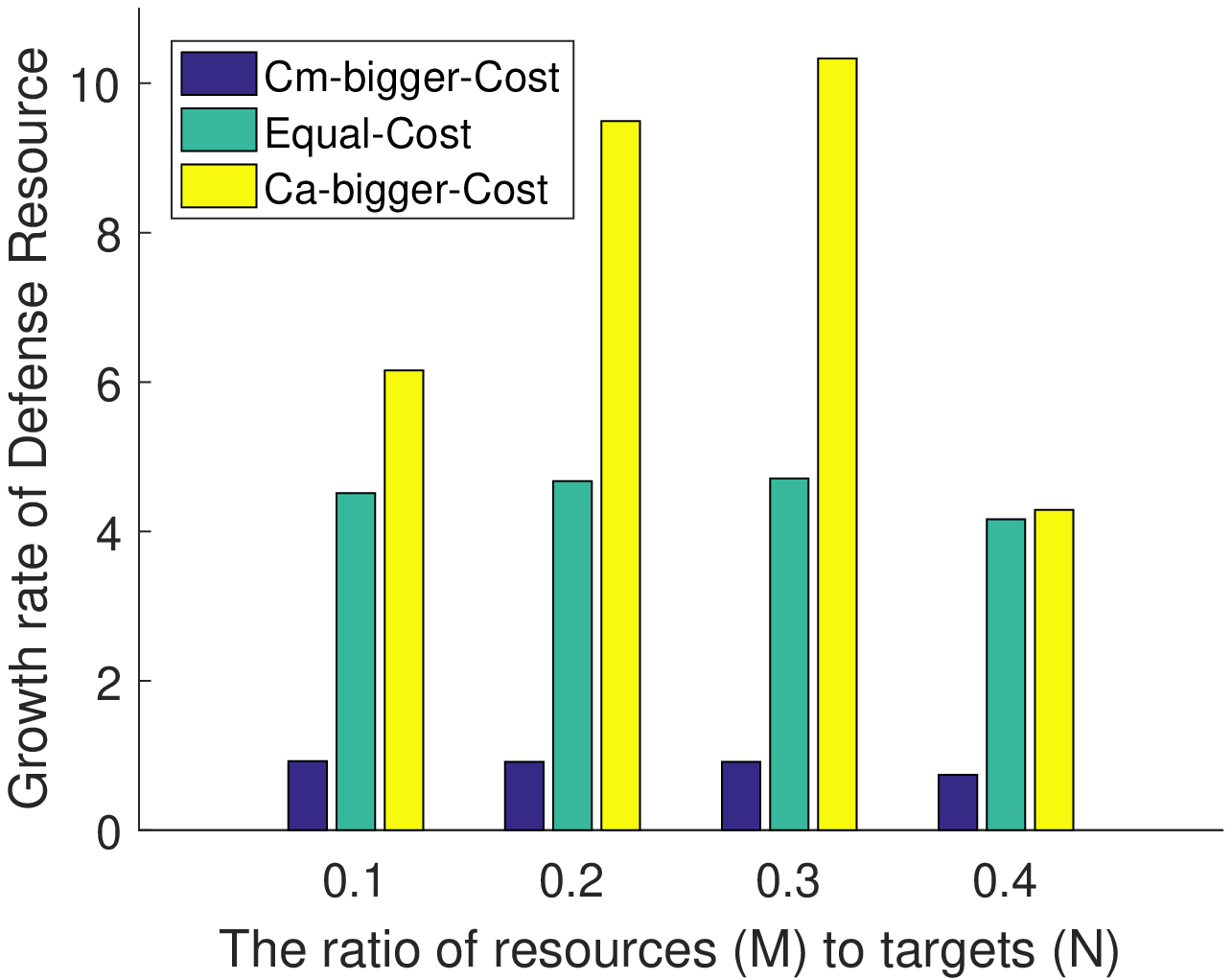}
\label{subfig:Re_rG}}
\hfil
\subfigure[Effectiveness comparison (ratio = 0.1)]{\includegraphics[scale=0.25]{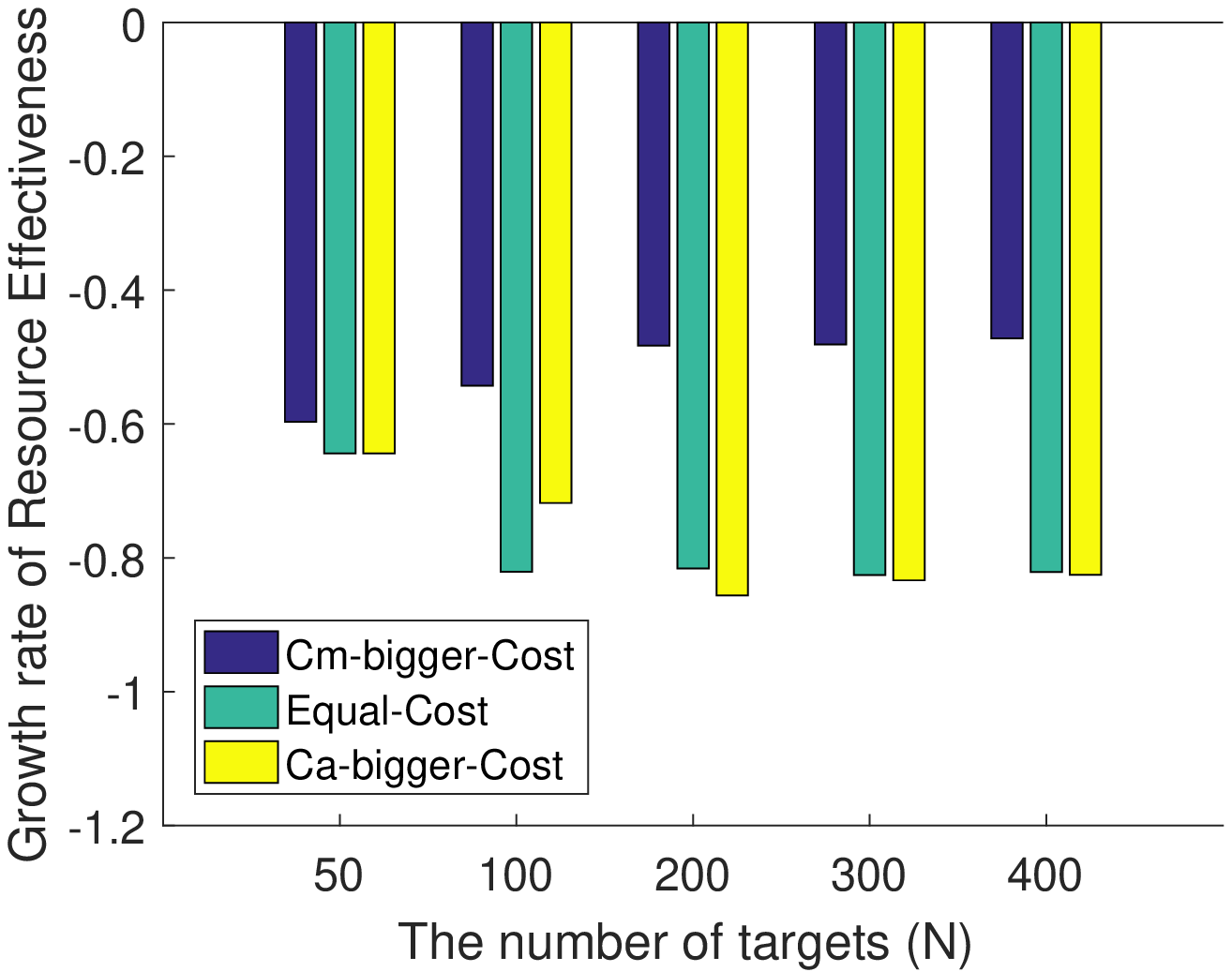}
\label{subfig:Effect_nG}}
\hfil
\subfigure[Effectiveness comparison (N = 200)]{\includegraphics[scale=0.25]
{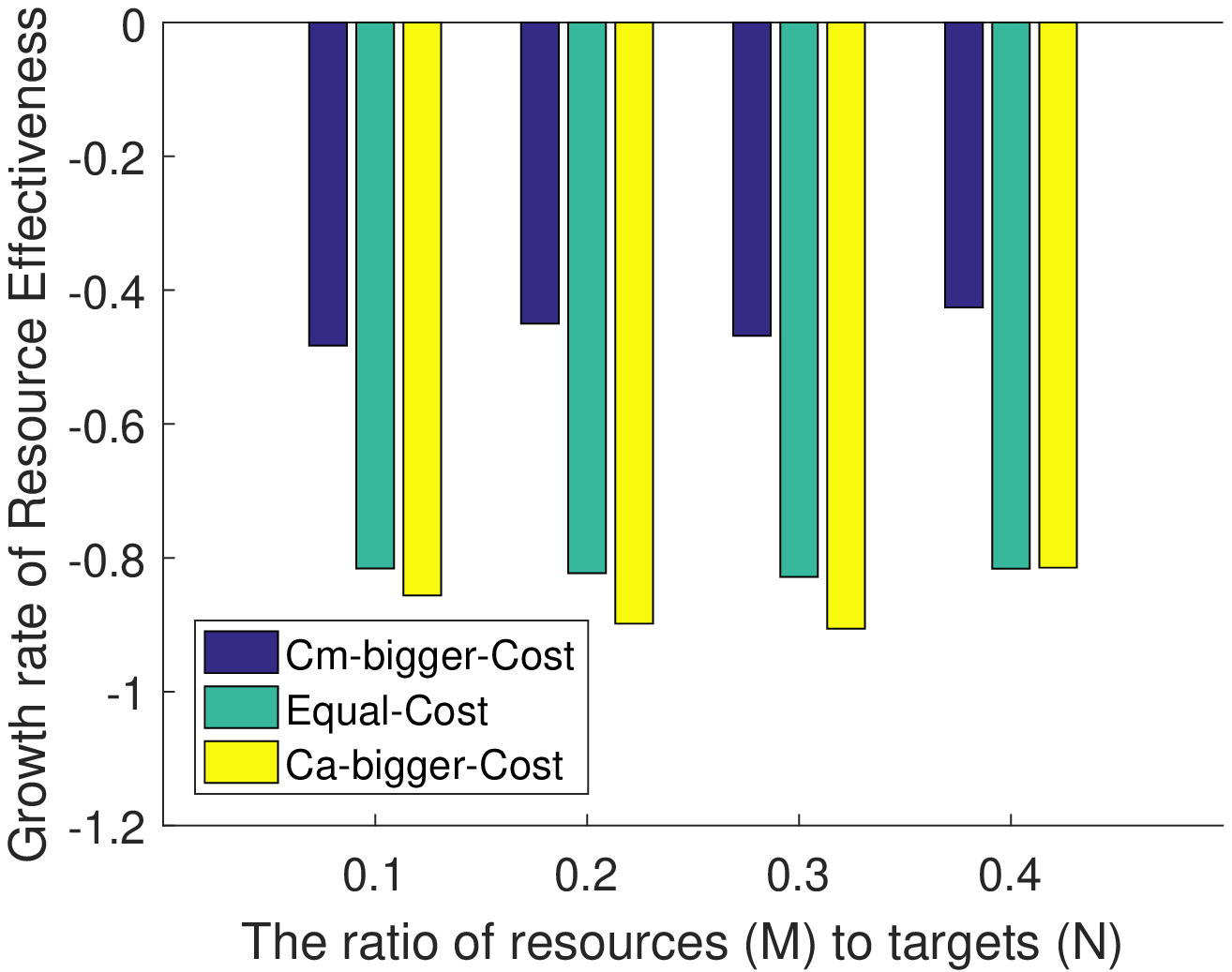}
\label{subfig:Effect_rG}}
\caption{\textbf{Difference in solution quality when the utility function does not consider the action cost.}}
\label{fig:SolutionGrowth}
\end{figure}

Furthermore, we illustrate the difference in solution quality when the utility function does not consider the action cost in Fig. \ref{fig:SolutionGrowth}. Equation \eqref{equ:effGrowth} shows that if the growth rate is less than zero, then $a$ is less than $b$. An overall comparative analysis of the results shown in Fig. \ref{fig:SolutionGrowth} indicates that when the utility function does not include the action cost, the defender's utility and the effectiveness are lower, and the defender's resource consumption is larger. The difference becomes especially evident as the number of resources increases (the resource-to-target ratio increases) or the number of targets increases ($N$ increases).

Overall, a utility function that considers the action cost, regardless of the relationship between defense and attack costs, provides the defender with a larger payoff and higher effectiveness. Additionally, although it is possible to represent the resources allocated to targets using the defense probabilities, as done in the utility functions implemented in many previous studies, sometimes the resource metric is not the same as the defense probability. For example, when 10 GB of storage is required to run intrusion prevention servers to protect a base station, this requirement cannot be represented as a probability. However, we can set $C_i^m = 10$ directly, and the defense probability then represents the probability that this target may be covered by this server. Hence, adding the action cost to the utility function is beneficial. In the following sections, we present a series of comparative analyses of various strategies based on our proposed utility function that considers the action cost.

\subsection{Comparison of Allocation Strategies}
A system with high security requirements is considered; e.g., government systems usually require a high level of consistency and need to be able to resist various attacks. The defender is usually equipped with high-performance defense modules with powerful processing capabilities, so a relatively large protection reward and a small protection penalty, which can be represented as $P_i^a >R_i^a $ \cite{reward}, are chosen in our study. Since all three scenarios regarding the relationship between the defense cost and the attack cost have a similar impact on the solution quality, we perform our further study based on the case in which the defense cost is less than the attack cost represented by $C_i^m < C_i^a$. 

We varied the reward and penalty from 1 to 10, the action cost from 0.1 to 0.4, and the numerical gap between the reward (or penalty) and the cost was considered large. Here, we limit the gap between the reward (or penalty) and the cost by randomly choosing values from $C_i^m \in [0.01,0.02]$, $C_i^a \in [0.02,0.03]$, $P_i^a \in [1.4,1.6]$, and $P_i^m \in [0.4,0.6]$.  These digits can be projected to the scenario that a unit of defense resource can protect at most 100 targets, and a unit of attack resource can attack at most 50 targets. If an attack fails, the attacker will get a penalty about 1.4. And if the protection fails, the defender will get a penalty about 0.4. In this case, the attacker can be seen as the type of risk-averse player who aims to minimize the risk loss \cite{Risk2017}.

To further evaluate the utility function which includes the action cost, we simulate four extreme resource allocation strategies in which the defender does not follow the NE strategy.

\subsubsection{PartOneS strategy}
The defender cannot protect all the targets due to resource limitations, so it must select at most $k$ targets to protect. The remaining $N-k$ targets are not protected. The defense probability distribution is obtained from \eqref{partones}. In this strategy, $M$ available units of resources are consumed.

\begin{eqnarray}
\scalebox{0.9}{$ q_i = \left\{ \begin{array}{l}
1,\quad i = 1,...,k - 1;\\
({M} - \sum\nolimits_{j = 1}^{k - 1}{q_j * C_j^m})/{C_i^m} ,\quad q_j=1,\quad i = k; \\
0,\quad i = k + 1,...,n.
\end{array} \right.  $}
\label{partones}
\end{eqnarray}

\subsubsection{Rand strategy}
The defender protects targets following a random probability distribution according to \eqref{rand}. In this strategy, the quantity of resources consumed is less than $M$.

\begin{equation}
\scalebox{1}{$ q_i =  \frac{Rand({q_i})*{M}}{\sum\nolimits_{j = 1}^n {Rand({q_j})*C_j^m}} ,  i=1,2,...,n $}
\label{rand}
\end{equation}

\subsubsection{Average strategy}
Resources are allocated to each target equally; the defense probability distribution obeys \eqref{average}. In this strategy, all $M$ available units of resources are consumed.

\begin{eqnarray}
 q_i*C_i^m={M}/{n},\quad i=1,2,...,n 
\label{average}
\end{eqnarray}

\subsubsection{AllOneS strategy}
The resource limitation is relaxed, and the defender protects all targets, as expressed in \eqref{allones}, which is approximately abstracted as \textit{AllOneS}. In this strategy, the quantity of resources consumed is greater than $M$.

\begin{eqnarray}
 q_i=1,\quad i=1,2,...,n 
\label{allones}
\end{eqnarray}

Our strategy obtained based on the proposed GTRA model is similar to \eqref{our}. It is computed using the IGA given in Algorithm \ref{algorithm1}. In our strategy, the quantity of resources consumed is no more than $M$.

\begin{eqnarray}
 \textit{\textbf{q}}=IGA(U_M,n,M), \quad   \sum q_i * C_i^m \leq M
 \label{our}
\end{eqnarray}

We start by comparing the utility of the defender with that of the attacker. The defender's resources are considered to be limited, whereas the attacker's resources are unlimited. One hundred game instances, with the number of targets ranging from 10 to 1000 in increments of 10, are considered.

\begin{figure}[!]
\centering
\includegraphics[scale = 0.35]{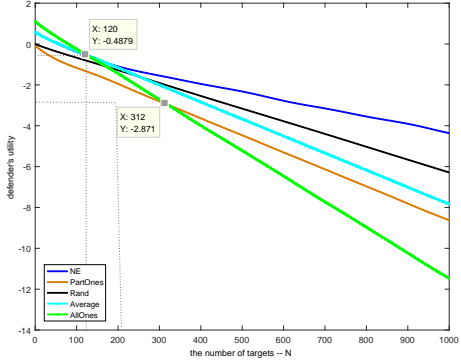}
\caption{\textbf{Comparison of the defender's utility.}\label{fig4}}
\end{figure}

The defender's utility results for the different strategies are displayed in Fig. \ref{fig4}. The vertical axis represents the defender's utility $U_M$, and the horizontal axis shows the number of targets $N$.

When the number of targets is below 120, \textit{AllOneS} provides the defender with the greatest utility. However, when the number of targets exceeds 312, \textit{AllOneS} provides the defender with the smallest utility because of the higher resource consumption. The NE strategy based on our GTRA model outperforms the other four strategies when the number of targets is greater than 120. 

Interestingly, the defender's utility decreases with an increasing number of targets in Fig. \ref{fig4}, while the opposite trend is seen in Fig. \ref{fig:Solution}. The difference between these two configurations is the range of parameters. When the reward or penalty is much larger than the cost, the defender's utility is larger, and the impact of the number of targets is directly proportional to the utility. By contrast,  when the reward or penalty is only slightly larger than the cost, the defender's utility is smaller and potentially even negative. In this case, the impact of the number of targets is directly proportional to the utility. Hence, the parameter configuration, such as the gap between the reward (or penalty) and cost, influences the defender's utility. Regardless of the parameter configuration, the NE strategy based on our GTRA model is better than the other strategies in terms of the defender's utility.

\subsection{Comparison in Terms of Various Evaluation Criteria}
The NE strategy is obtained by finding the maximum utility for both players. In this subsection, we evaluate the vulnerability, coverage and effectiveness of our equilibrium strategy and the other four strategies.

\subsubsection{Vulnerability}
We first evaluate the vulnerability of the defender's various strategies. The $vulnerability$ is defined as a risk indicator for the targets as shown in \eqref{equ:vul} \cite{vul}.

\begin{eqnarray}
vulnerability = \frac{success-failure}{success+failure}
\label{equ:vul}
\end{eqnarray}

\noindent where $success$ and $failure$ denote the numbers of targets in which an attack that is launched is not detected or is detected, respectively. The assumption is made that if the defender allocates resources to protect a target, then that target will be successfully protected against attack by the continuously operating defense system; otherwise, the attack will be successful. Clearly, a greater $success$ value and a lower $failure$ value indicate a more vulnerable strategy. Hence, the lower the $vulnerability$ is, the better the strategy.

\begin{figure}[!]
\centering
\includegraphics[scale = 0.35]{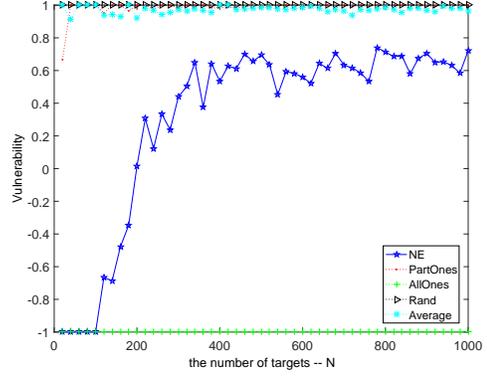}
\caption{\textbf{Vulnerability of 100 groups of instances.}\label{fig6}}
\end{figure}

Fig. \ref{fig6} shows that the $vulnerability$ of \textit{AllOneS} is $-1$, which implies that the number of successes is $0$. In this situation, the targets are the most secure. The defender's protections cover all the targets, resulting in the most secure environment. When the number of targets (\textit{N}) is small, the \textit{NE} strategy achieves the most secure state; as \textit{N} increases, the $vulnerability$ increases because the available resources become insufficient to protect all targets. Additionally, once \textit{N} is greater than 400, the $vulnerability$ of \textit{NE} varies only slightly. These analyses reveal that \textit{NE} can be scaled up to protect a large number of targets. Furthermore, compared with \textit{PartOneS}, \textit{Rand}, and \textit{Average}, as the number of targets to protect increases such that there are insufficient resources to protect all of them, \textit{NE} performs better. It enables control of the trade-off between the security benefit and the resource consumption and focuses on protecting targets that are more likely to be attacked. 

As a result, \textit{NE} performs better than all the other strategies except for \textit{AllOneS} in terms of the $vulnerability$ of the targets.

\subsubsection{Coverage}
We evaluate the allocated resources' coverage of the targets next. The \textit{coverage} is defined as the proportion of protected targets among the total targets, as shown in \eqref{equ:protec}, where the protected targets are defined as those that are attacked by attacker and also protected by defender, those that are not attacked but protected, and those that are not attacked and not protected, either, as denoted by \textit{AP}, \textit{NP} and \textit{NF}, respectively, in Table \ref{tab4}.

\begin{table}[h]
\centering
\caption{\bf Protection type of target $i$}
\begin{tabular}{|c|c|c|}
\hline
\multicolumn{1}{|l|}{\bf ~ } & \multicolumn{1}{|l|}{\bf Protect}& \multicolumn{1}{|l|}{\bf Fail to Protect} \\ 
\hline
Attack & AP & AF \\
\hline
Not Attack & NP & NF\\
\hline
\end{tabular}
\label{tab4}
\end{table}
\begin{eqnarray}
coverage = \frac{AP + NP + NF}{N}
\label{equ:protec}
\end{eqnarray}

\begin{figure}[!]
\centering
\includegraphics[scale = 0.35]{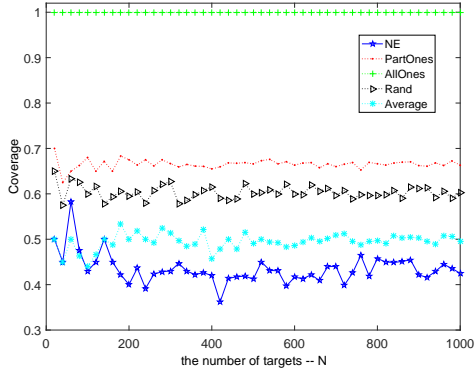}
\caption{\textbf{Coverage in 100 groups of instances.}\label{fig7}}
\end{figure}

Fig. \ref{fig7} shows the coverage results for the five strategies for 100 groups of experimental instances. \emph{AllOneS} covers all targets by protecting all of them, whereas \emph{NE} covers almost the fewest targets because \emph{NE} protects only risky targets attractive to the attacker. Thus, the number of protected targets is smaller than the other strategies.The disadvantage of this strategy is that it cannot guarantee the absolute security of the targets, in contrast to \emph{AllOneS}. However, it may be useful for saving resources or improving the effectiveness with which those resources are used, especially when resources are valuable or limited.

\subsubsection{Effectiveness}
\begin{figure}[!]
\centering
\subfigure[effectiveness.]{\includegraphics[scale=0.33]
{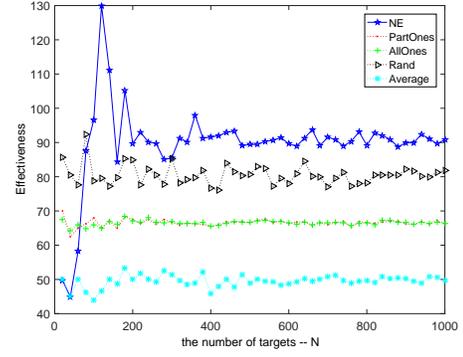}
\label{subfig:Effec}}
\hfil
\subfigure[Defender's resource consumption.]{\includegraphics[scale=0.445]{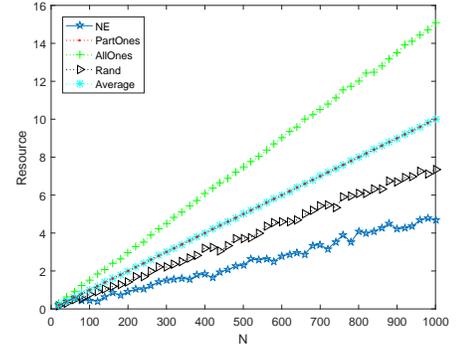}
\label{subfig:Res}}
\caption{\textbf{Defender's effectiveness.}}
\label{fig:ReEffect}
\end{figure}

We now evaluate the effectiveness of the five strategies. The greater the effectiveness is, the better the strategy is. For 100 groups of experimental instances, the quantities of resources consumed by each strategy are plotted in Fig. \ref{subfig:Res}. It is worth noting that \emph{AllOneS} consumes the most resources, and \emph{NE} consumes the least resources. The  resources consumed by \textit{PartOneS} and \textit{Average} are equivalent since these two strategies use all the available resources. When the number of targets is increased to 1000, the quantity of resources consumed by \textit{AllOneS} is close to four times the resources consumed by \textit{NE}. When these values are applied to the real world, they represent a large amount of material or financial resources that must be expended by the defender. Consequently, our strategy aims to provide high effectiveness.

In Fig. \ref{subfig:Effec}, there is an evident upward trend in the effectiveness of \emph{NE} when the number of targets is less than 200, which then gradually drops to a stable value with an increasing number of targets. Moreover, \emph{NE} has the highest effectiveness among all five strategies. These results suggest that increasing the number of targets does not affect the effectiveness. In addition, although \textit{AllOneS} protects the most targets, its effectiveness is lower than that of \textit{NE} because it consumes more resources. \textit{AllOneS} may protect some targets that are not likely to be attacked, which may cause resources to be consumed without gaining benefits, thus decreasing the defender's effectiveness. 

\begin{figure}[!]
\centering
\subfigure[N vs. coverage vs. vulnerability.]{\includegraphics[scale=0.35]{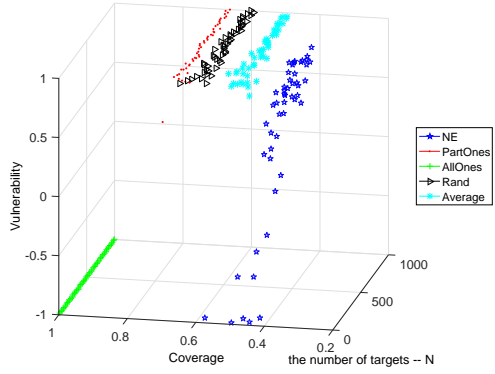}
\label{subfig:NCV}}
\hfil
\subfigure[N vs. effectiveness vs. vulnerability.]{\includegraphics[scale=0.35]
{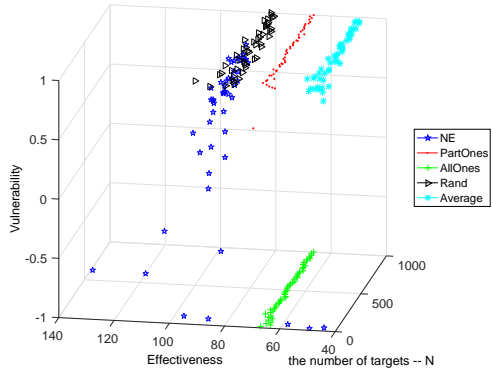}
\label{subfig:NEV}}
\caption{\textbf{Comprehensive analysis of the number of targets, coverage, vulnerability and effectiveness.}}
\label{fig:NCEV}
\end{figure}

Now, we combine the number of targets, coverage and vulnerability in Fig. \ref{subfig:NCV} and combine the number of targets, effectiveness and vulnerability in Fig. \ref{subfig:NEV}. From Fig. \ref{fig:NCEV}, it can be concluded that more targets must be protected to maintain a low vulnerability or to decrease the vulnerability. However, if the defender increases the number of protected targets, more resources will be required. Take \textit{AllOneS} as an example. The vulnerability of the targets is near zero, and the number of protected targets is the largest, but the number of covered targets per resource is low because of the high resource consumption. In this situation, to improve the security of the targets, the NE strategy obtained based on the proposed \textit{NE} model, which balances the security utility and the resource consumption, is the best choice for allowing the defender to utilize limited resources effectively.

\subsection{Parameter Analysis}
The security utilities of the defender and the attacker are related not only to their strategies but also to certain specific parameters: the resource constraint $M$, the prediction accuracy $\alpha$ and the noise $\lambda$ in the attacker's rationality.

\subsubsection{Resource constraint $M$}
\begin{figure}[!]
\centering
\subfigure[Defender's utility]{\includegraphics[scale=0.45]{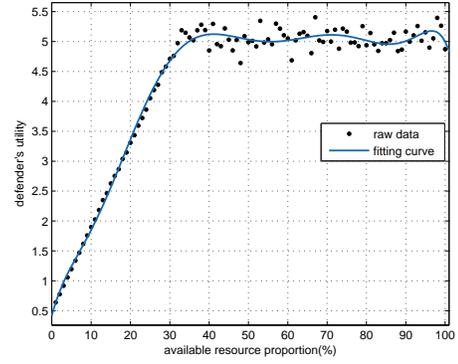}
\label{subfig:Um_M}}
\hfil
\subfigure[Attacker's utility]{\includegraphics[scale=0.45]
{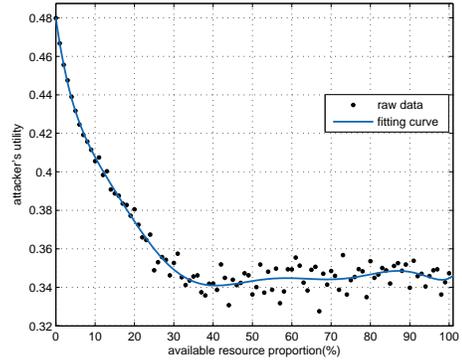}
\label{subfig:Ua_M}}
\caption{\textbf{Impact of the quantity of resources on the players' utilities.}}
\label{fig:DifM}
\end{figure}

We generate 100 random game instances with 1000 targets and consider different quantities of resources to assess the impact of the resource constraint $M$ on the players' utilities. In Fig. \ref{fig:DifM}, the x-axes show the proportion of available resources relative to the maximum resources required, and the y-axes represent the players' utilities. 

Fig. \ref{fig:DifM} shows that when the resource proportion is zero, the defender's utility is the lowest, and the attacker's utility is the highest. As the proportion of available resources increases, the defender's utility increases, and the attacker's utility decreases. When the proportion reaches 40\%, the utilities of both players become stable. Hence, we conclude that for the case in which $R_i^m > P_i^m$ and $P_i^m > C_i^m$, 40\% of the maximum resources is an efficient rate of utilization for the defender. When the proportion is greater than 40\%, both players' utilities remain approximately stable. The jitter in the raw data is due to the aggregated analysis of resource consumption, which demonstrates that spending more resources to protect targets may be less risky, but the cost of the resources consumed will exceed the benefit. 

The proposed \textit{NE} model can compute the corresponding best resource proportions for different combinations of reward, penalty and cost. Therefore, the proposed model offers the defender an alternative means of gaining greater utility while saving resources, thereby improving the defender's outcome from the perspective of economics. When the defender needs to estimate the overall quantity of resources required to protect a massive number of targets, the proposed GTRA model can be used to compute the approximate quantity based on the configurations of all the targets and thus provide the defender with a game theoretical reference value.

\subsubsection{Prediction accuracy $\alpha$}
\begin{figure}[!]
\centering
\subfigure[Defender's utility]{\includegraphics[scale=0.4]{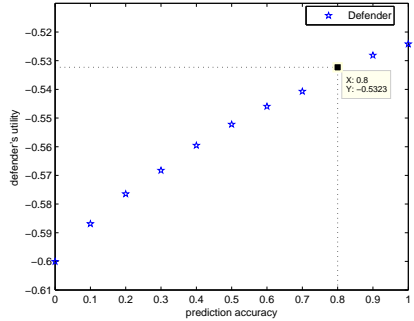}
\label{subfig:Um_a}}
\hfil
\subfigure[Attacker's utility]{\includegraphics[scale=0.4]
{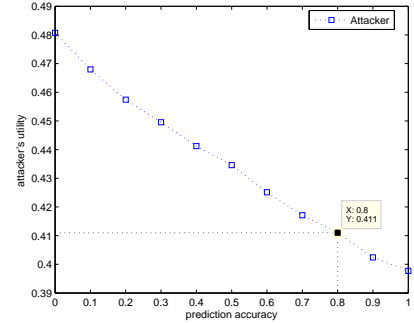}
\label{subfig:Ua_a}}
\caption{\textbf{Impact of $\alpha$ on the players' utilities.}}
\label{fig:Difa}
\end{figure}

We generate 10 random game instances with 100 targets and vary the prediction accuracy $\alpha$ to assess its impact on the players' utilities. $\alpha$ is the accuracy with which attacks are predicted by the defender. Figs. \ref{subfig:Um_a} and \ref{subfig:Ua_a} show the differences in the players' utilities with varying $\alpha$ values (ranging from 0 to 1). The prediction accuracy $\alpha$ is plotted on the horizontal axis, and the player's utility is plotted on the vertical axis.

As the prediction accuracy increases, the defender's utility increases, and the attacker's utility decreases. For a typical state in which $\alpha$ is 0.8, the defender's utility is -0.5323, and the attacker's utility is 0.411. The reason that the sum of the defender's utility and the attacker's utility is not equal to zero is that our game is a non-zero-sum game. In this paper, we assume that the predictions are not fully accurate, so we take $\alpha$ to be 0.8 without explicit explanation.

\subsubsection{Noise $\lambda$ in the attacker's rationality}

\begin{figure}[!h]
\centering
\includegraphics[scale = 0.5]{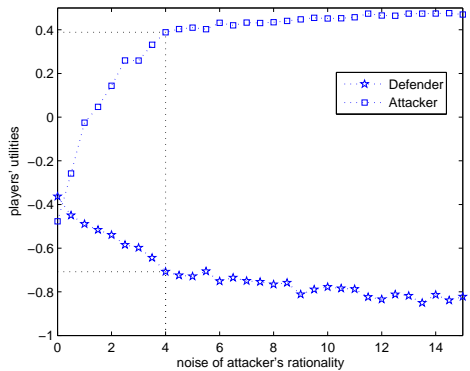}
\caption{\textbf{Impact of $\lambda$ on both players' utilities.}\label{fig16}}
\end{figure}

We generate 30 random game instances with 100 targets and vary $\lambda$ to assess its impact on the players' utilities. $\lambda$ represents the noise in the attacker's rationality during strategy planning. We vary $\lambda$ from 0 to 15 in increments of 0.5. In Fig. \ref{fig16}, the two variables are the noise $\lambda$ in the attacker's rationality and the player's utility. $\lambda$ is the independent variable, and the utility is the dependent variable. The change in the utility is caused by different values of $\lambda$.

Fig. \ref{fig16} shows that the larger $\lambda$ is, the greater the utility of the attacker is and the lower the utility of the defender is. Additionally, when $\lambda$ is greater than 4, both players' utilities remain nearly stable with $\lambda$ increasing, especially that of the attacker. We may argue that if the attacker is sufficiently rational ($\lambda$ is sufficiently high), then the players' utilities are nearly constant. In this paper, to model irrational attack behavior, the value of $\lambda$ is set to 1.5 in the analysis \cite{27}.

\section{Conclusion}\label{sec:conclusion}
\subsection{Summary}
In this paper, we investigate how to allocate resources to efficiently protect targets when the number of targets is greater than the number of resources. A game theoretic resource allocation (GTRA) model is constructed based on a Stackelberg game. In the proposed model, an independent item (i.e., the action cost) is included in the game utility function compared with the previous studies, which makes the resource allocation more flexible and convenient. The proposed method correlates resource allocation with security by means of the game utility function, simulates the behavior of an attacker of the adversarial nature through the introduction of the QR model, and enables the computation of the Nash equilibrium (NE) strategy through an iterative genetic algorithm.

In addressing these challenges, we draw the following conclusions:

\begin{itemize}
\item Including the action cost in the utility function provides the defender with greater utility and higher effectiveness, regardless of the relationship between the defense cost and the attack cost.

\item The size of the gap between different parameters affects the defender's utility and the trend of variation in the defender's utility with the number of targets. Regardless of the parameter configuration, the NE strategy based on our GTRA model outperforms the other four resource allocation strategies considered for comparison.

\item When the available resources are not sufficient to protect all the targets, our strategy performs better than the random allocation strategy, the average allocation strategy and the partial protection strategy. It can effectively balance security and resource consumption.

\item When the resource constraint is relaxed, although our strategy cannot maintain the best target security, it nevertheless achieves higher effectiveness than the one allocating resources to all the targets. Thus, it can optimize the consumption of resources for protecting targets.

\item The quantity of resources and the security of the targets are not directly related. Given a set of targets and their corresponding asset values, the proposed model provides advice on the quantity of resources required to effectively protect the targets. 
\end{itemize}

Given these findings, the security of targets can be better protected by considering the cost of protection when planning resource allocation. Last but not least, we hope that this study can serve as a theoretical reference for the allocation of security resources in multiple arenas. 

\subsection{Future work}
Our current work focuses on designing an efficient resource allocation strategy to protect a massive number of targets using limited resources. Next, we plan to apply current research in an application leveraging the idea of software-defined networking (SDN) and network function virtualization (NFV), which is suitable not only for common networks but also for computing environments such as cloud computing.

\section*{Conflict of interest}
The authors declare that they have no conflicts of interest.

\section*{Acknowledgments}
This work was partially supported by the Science and Technology Development Plan Projects (20150204081GX and 20180414024GH) of Jilin Province of China, and the 13th Five-Year Science and Technology Research Project of the Education Department of Jilin Province under Grant No. JJKH20190598KJ.
.

\bibliography{MyRef}

\end{document}